\documentclass[11 pt, a4paper]{article}
\usepackage{array,epsfig}
\usepackage{amsthm}
\usepackage{amsmath}
\usepackage{amssymb}
\allowdisplaybreaks 
\usepackage{color}
\usepackage{enumitem}
\usepackage{booktabs}
\usepackage{sidecap}
\usepackage{setspace}
\usepackage{comment}
\usepackage{url}
\usepackage{float}
\usepackage[normalem]{ulem}
\usepackage{apacite} 
\usepackage{parskip}
\usepackage{standalone}
\usepackage{tabu}
\usepackage[font=footnotesize,labelfont=bf]{caption}

\usepackage{bm, bbm, mdframed, mathtools, mathabx, multirow, subfigure, tikz, makecell, tcolorbox, soul, caption}
\usepackage[left=2.5cm, right=2.5cm, top=3cm, bottom=3cm]{geometry}
\usepackage[english]{babel}
\usepackage[mathscr]{euscript}
\newcommand{\E}{\mathrm{E}}

\numberwithin{equation}{section} 
\tcbuselibrary{breakable}
\usepackage{mathdots}
\usepackage{lscape}
\usepackage{pdflscape}

\usepackage[round, longnamesfirst]{natbib}
\usepackage[T1]{fontenc} 
\usepackage[autostyle]{csquotes}  
\usepackage[citecolor=black, linkcolor = black, colorlinks=true, allcolors=black]{hyperref}
\usepackage{glossaries}
\usepackage{appendix}
\usepackage{verbatim}
\usepackage{tcolorbox}
\usepackage{xcolor}
\usepackage{listings}
\usepackage{rotating}
\usepackage{comment}
\usepackage{dsfont}
\usepackage{algorithm2e}
\usepackage{bookmark}
\usepackage{setspace}
\usepackage{xr}
\usepackage{threeparttable}
\usepackage{subcaption}
\usepackage{graphicx}

\RestyleAlgo{ruled}
\SetKwComment{Comment}{/* }{ */}
\usepackage{tikz}
\usetikzlibrary{shapes.geometric, arrows, positioning, arrows.meta}
\tikzstyle{c_solid} = [circle, minimum width=1cm, minimum height=1cm,text centered,draw=black, fill=white]
\tikzstyle{c_dashed} = [circle, minimum width=1cm, minimum height=1cm,text centered,draw=black, dashed]
\tikzstyle{arrow} = [thick,->,>={Stealth[scale=1.3]}]
\newtheorem{lemma}{Lemma}
\newtheorem{definition}{Definition}

\newcommand{\ind}{\perp \!\!\!\! \perp}



\makeatletter
\renewenvironment{abstract}{%
  \if@twocolumn
    \section*{\abstractname}%
  \else
    \small
    \begin{center}%
      {\bfseries \abstractname\par}%
    \end{center}%
    \noindent\ignorespaces
  \fi
}{\par\unskip}
\makeatother

\title{\textbf{Fairness-Aware and Interpretable Policy Learning}}
\author{
  Nora Bearth\thanks{University of St.Gallen, Rosenbergstrasse 22, 9000 St.Gallen, CH, E-mail: \texttt{nora.bearth@unisg.ch}, \texttt{michael.lechner@unisg.ch}, \texttt{jana.mareckova@unisg.ch} (corresponding author), \texttt{fabian.muny@unisg.ch}\\
   }
    , Michael Lechner\textsuperscript{\footnotemark[1]}
\textsuperscript{\footnotemark[2]}\thanks{Michael Lechner is also affiliated with Örebro University, CEPR, London, CESIfo, Munich, IAB, Nuremberg and IZA, Bonn. \\
Financial support from the Swiss National Science Foundation (SNSF) is gratefully acknowledged. The study is part of the project "Chances and risks of data-driven decision making for labour market policy" (grant number SNSF 407740\_187301) of the Swiss National Research programme "Digital Transformation" (NRP 77). Editing was supported by GPT-4 and Grammarly.}
\addtocounter{footnote}{1},
Jana Mareckova\textsuperscript{\footnotemark[1]},
    Fabian Muny\textsuperscript{\footnotemark[1]}
}

\date{}
\begin{document}
\begingroup
\let\newpage\relax
\maketitle
\endgroup

\begin{center}
\textit{This version: \today} \\
Comments welcome.
\end{center}

\vspace{0.5cm}
\begin{abstract}
Fairness and interpretability play an important role in the adoption of decision-making algorithms across many application domains. These requirements are intended to avoid undesirable group differences and to alleviate concerns related to transparency. This paper proposes a framework that integrates fairness and interpretability into algorithmic decision making by combining data transformation with policy trees, a class of interpretable policy functions. The approach is based on pre-processing the data to remove dependencies between sensitive attributes and decision-relevant features, followed by a tree-based optimization to obtain the policy. Since data pre-processing compromises interpretability, an additional transformation maps the parameters of the resulting tree back to the original feature space. This procedure enhances fairness by yielding policy allocations that are pairwise independent of sensitive attributes, without sacrificing interpretability. Using administrative data from Switzerland to analyze the allocation of unemployed individuals to active labor market programs (ALMP), the framework is shown to perform well in a realistic policy setting. Effects of integrating fairness and interpretability constraints are measured through the change in expected employment outcomes. The results indicate that, for this particular application, fairness can be substantially improved at relatively low cost.
\end{abstract}

\vspace{10pt}
{\small
\textbf{JEL classification:} C14, C21 \\
\textbf{Keywords:} Algorithmic decision making, causal machine learning, fair machine learning, treatment effect heterogeneity
}

\thispagestyle{empty}
\newpage

\setcounter{page}{1}

\section{Introduction} \label{2_introduction}

The rise of artificial intelligence and the growing availability of large-scale data have made algorithmic decision making increasingly common across sectors such as healthcare, education, finance, and public policy. Algorithms are often valued for their potential to enhance efficiency, uncover complex patterns, and reduce subjective human biases. However, significant concerns about potential algorithmic bias remain. A widely discussed example is the investigation of the COMPAS risk assessment tool conducted by the nonprofit newsroom ProPublica, which examines differences in recidivism risk scores across defendants grouped by race \citep[see, e.g.,][]{Propublica:2016,Flores:2016,Dressel:2018, Washington:2018}. When such scores are further used for decision making, ensuring the absence of bias is important, as they could otherwise negatively influence the final outcomes. 

Among various approaches to algorithmic decision making, policy learning has emerged as a particularly useful framework for design and evaluation of economic policies. Policy learning refers to the use of data to construct policies that optimize a specified objective function, such as a measure of social welfare.\footnote{Throughout this paper, \textit{policy} is used as a synonym for \textit{decision rule}. An \textit{assigned treatment}, an \textit{allocation} and a \textit{decision} denote the result of applying the policy to specific units.}
Unlike standard classification tasks, which aim to replicate historical patterns, policy learning follows a prescriptive approach. It assigns treatments based on units' features with the goal of optimizing the chosen objective, often defined as the expected outcome of the assigned treatments.
As a result, a learned policy discriminates between units based on their observed features to make tailored treatment assignments. Although this form of statistical discrimination is necessary for effective decision making, it may lead to systematic disparities between groups, some of which the decision maker prefers to avoid.

To illustrate, consider a setting where the goal is to assign unemployed individuals to one of two active labor market programs, namely computer courses and vocational training. Assume that women benefit more from vocational training while men gain more from computer courses. Hence, optimizing solely for expected employment outcomes will result in assigning all women to vocational training and all men to computer courses. This allocation may be considered unfair by the decision maker if the aim is to ensure equal access to programs across the two groups. In general, even when a learned policy is optimal with respect to the decision-maker's main outcome criteria, it may still exhibit undesirable properties. In particular, it may systematically favor or disadvantage individuals based on attributes such as gender or race, which are often considered legally, socially or ethically inappropriate for decision making. 
A naive approach to mitigate such disparities is to restrict the model to features deemed relevant for decision making while excluding such sensitive attributes. However, if the features used for decision making correlate with the sensitive attributes, fairness concerns, such as those described above, will persist.

Deploying algorithmic decision making in the presence of sensitive attributes poses significant challenges in economic and social policy contexts, as outcomes perceived as unfair can  undermine trust in the algorithms and the responsible institutions.
Beyond fairness concerns, the lack of interpretability is another major concern when adopting algorithmic decision tools in practice. While interpretability is essential for a better understanding and oversight of algorithmic outputs, it does not, of course, guarantee fairness on its own. Addressing both concerns requires methodological approaches that promote fairness without compromising interpretability.

This paper proposes a method that promotes fairness while preserving interpretability in policy learning. Thus, it contributes to the ongoing discourse on ethical and transparent algorithmic decision making, offering practical insights for decision makers and researchers alike. It introduces an approach that integrates a common notion of fairness in algorithmic decision making, defined as statistical independence between assigned treatments and sensitive attributes. This independence is targeted via a pre-processing step that adjusts the data before applying a policy tree \citep{Athey:2021,Zhou:2022}. However, the pre-processing step may distort the interpretability of the resulting model. To address this, a procedure is proposed that modifies the policy tree to allow describing the estimated policy in terms of the original features while preserving fairness.
Indeed, the method extends to threshold-based policies, beyond trees.

The procedure is applied to a real-world dataset to demonstrate its practical applicability. The empirical analysis uses administrative data from Switzerland to study the allocation of unemployed individuals to different active labor market programs and examines the 
interplay between fairness, interpretability, and efficiency, measured in terms of average employment chances. 
The empirical results suggest that in this particular setting the costs of incorporating interpretability and fairness are relatively low, with fairness constraints leading to only minor reductions in average employment chances compared to the fairness-unaware interpretable allocation. However, when comparing the fairness-unaware interpretable policy to the fairness-aware interpretable policy, reallocations across programs, which are needed to achieve fairness, result in both gains and losses for different types of unemployed. Importantly, these changes occur mainly within groups with weaker labor market attachment, indicating that the policy adjustment redistributes resources among members of this group rather than shifting them between individuals with weaker and stronger labor market attachment.

The paper is structured as follows: Section \ref{related_literature} reviews the relevant literature. Section \ref{Setting} introduces the notation and outlines the policy learning setting. Section \ref{sec:pre_poc} presents the approach for incorporating fairness into policy learning, detailing both the pre-processing step and the adjustments made to the standard policy tree to retain interpretability. It also discusses practical solutions to challenges that arise when aligning fairness and interpretability. Section \ref{Empirical_example} illustrates the application of the framework in a real-world scenario, and Section \ref{conclusion} concludes. Additional information on the empirical study is provided in the Appendix \ref{Appendix_A}.

\section{Related Literature} \label{related_literature}

\subsection{Policy Learning}

\defcitealias{Chernozhukov:2018}{Chernozhukov, Chetverikov, Demirer, Duflo, Hansen, Newey, \& Robins  (2018)} 

Algorithmic decision making entered the econometrics literature through research on statistical treatment rules \citep{Manski:2004,Hirano:2009,Kitagawa:2018,Athey:2021,Zhou:2022}. The core idea is to estimate unit-specific scores measuring the value of each treatment alternative at a fine-grained aggregation level based on training data. These scores are then used to construct treatment assignment rules that map observable features to treatments in a way that maximizes a welfare criterion\footnote{In the policy learning context, the \textit{welfare criterion} is typically quantified as the \textit{policy value}, which often represents the expected outcome under a given policy.}, a process known as policy learning or empirical welfare maximization. The literature originates with \citet{Manski:2004}, who takes minimax regret as a criterion for evaluating statistical treatment rules. Building on this, \citet{Kitagawa:2018} develop an empirical welfare maximization approach using inverse-probability weighting for binary treatments. \citet{Athey:2021} extend this to observational settings with unknown treatment probabilities using a doubly robust learner with cross-fitting, as in \citetalias{Chernozhukov:2018}. \citet{Zhou:2022} further generalize the framework to multiple treatments and propose efficient numerical algorithms for policy trees, a class of decision trees specifically designed for policy learning.

\subsection{Transparency of Decisions}

Algorithmic decision systems offer new possibilities for data-driven decision making, often in collaboration with humans. However, prior research shows algorithmic aversion, where individuals tend to prefer human over algorithmic decisions \citep[e.g.][]{Burton:2020, Dietvorst:2015}. Key to encouraging the use and acceptance of algorithms is trust in the system \citep{Choung:2023} and users' understanding of the decisions communicated \citep[e.g.][]{Panigutti:2022, Bansal:2021}. \cite{Senoner:2024} find that transparent algorithms improve expert performance by increasing adherence to accurate algorithmic recommendations and facilitating rejection of inaccurate ones.
Transparent decision rules can be achieved in two ways. One is to directly use inherently \textit{interpretable models}, such as policy trees or rule-based learners (mentioned also in \cite{Athey:2021} or \cite{Zhou:2022}). Alternatively, post-hoc \textit{explanation methods}, like variable importance or partial dependence plots, can be applied to describe Blackbox models \citep[][Chapter 1]{Molnar:2020}. This study pursues the first approach, focusing on tree-based methods.

\subsection{Algorithmic Fairness}

Algorithmic fairness first emerged in the classification literature within computer science, addressing concerns over statistical disparities in predictions across groups. Early research emphasizes that machine learning algorithms could systematically disadvantage units based on so-called sensitive attributes (e.g.~\citet[][Chapter~3]{Barocas:2017}, ~\citet[][]{Feldman:2015}). To address these issues, various fairness criteria are proposed, some of which are mutually exclusive (see e.g.~\citet[][Chapter~3]{Barocas:2017} for an overview). Conceptually, fairness definitions can be categorized as either ``observational'' or causal, depending on whether they rely solely on observable variables or on modeled counterfactual relationships.\footnote{Note that the term ``observational'' in observational fairness differs from its conventional use in causal inference, where observational data may still permit causal analysis under suitable identification assumptions.} A widely used observational criterion is statistical parity,\footnote{Also known as demographic parity; its violation is referred to as disparate impact.} which requires predicted outcomes to be independent of sensitive attributes. This study adopts a version of statistical parity tailored to the policy learning context (see Section \ref{sec:fair_def}). A key advantage of observational fairness definitions is that they can be empirically validated using samples from the joint distribution of variables (subject to statistical sampling error).\footnote{Although policy learning can be formulated as a causal problem, this paper approaches fairness using observational rather than causal fairness definitions. In the policy learning setting, causal fairness \citep[e.g.][]{Kusner:2017, Kilbertus:2017, Nabi:2018, Chiappa:2019, Salimi:2019} would necessitate modeling the structural causal model between sensitive attributes and potential realizations of the features underlying the treatment assignment rule. This paper avoids imposing restrictive structural assumptions, at the cost of only addressing the observable aspects of unfairness rather than its fundamental causes.}

\defcitealias{Li:2022}{Li, Meng, Chen, Yu, Wu, Zhou, \& Xu  (2022)} 

Fairness requirements are typically implemented at one of three stages of the workflow: \textit{pre-processing}, which modifies training data before learning; \textit{in-processing}, which adjusts the algorithm itself; and \textit{post-processing}, which alters predictions after model training. For a detailed review, see \cite{Hort:2023}. This paper focuses on pre-processing approaches, which can be grouped into several categories. \textit{Relabelling} modifies outcomes for certain observations to balance predictions between groups. For example, \citet{Kamiran:2012} and  \citet{Kamiran:2013} predict outcomes using all features, including sensitive attributes, and then relabel units near the decision boundary to equalize outcomes across sensitive groups. A classifier is then retrained using only non-sensitive features and the adjusted labels. \textit{Perturbation} methods modify variables to align the distributions of sensitive groups while preserving within-group ranks. \cite{Feldman:2015} implement this strategy in a univariate setting and suggest applying it separately to each variable in multivariate cases. \cite{Johndrow:2019} extend this to multivariate settings using sequential transformations to achieve mutual independence from sensitive attributes. \cite*{Wang:2019} adjust distributions of the unfairly treated group to resemble the baseline, while \citetalias{Li:2022} residualize decision-relevant features to ensure mean-independence from sensitive attributes. \textit{Sampling} approaches adjust the sample distribution via reweighting, addition, or removal of observations \citep{Kamiran:2012}. Finally, \textit{representation learning} aims to map data into a space that fairly represents the underlying structure across groups \citep{Zemel:2013}. While distinct in their implementation, perturbation and representation learning methods share a common objective, as both aim to reduce the dependence of decision-relevant features on sensitive attributes while preserving as much information from the original features as possible.
The procedure proposed in this paper draws mainly on perturbation strategies, as detailed in Section \ref{sec:pre_poc}.

\subsection{Policy Learning and Fairness}

Fairness criteria from the classification literature can be reinterpreted in the context of policy learning. Following \citet{Frauen:2024}, there exist two perspectives on ``observational'' fairness in this setting. The first, \textit{action fairness}, demands that treatments recommended by the policy should be fair. 
Action fairness is violated when units from different sensitive groups have unequal probabilities of receiving a particular treatment. On the other hand, \textit{value fairness} requires that the distribution of outcomes resulting from the assigned treatments should be fair. Thus, even if a certain treatment is disproportionately assigned to a group with specific sensitive attributes, it would not be deemed unfair, provided that it results in more equal outcomes. The choice between fairness perspectives depends on the preferences of the decision maker. Action fairness could be an appropriate choice for decision making under the goal of achieving equal opportunity to a beneficial treatment, which is a relevant scenario in many applications. It is also often a requirement from a legal point of view not to discriminate against certain sensitive groups (e.g., Swiss Federal Constitution, Art.~8, para.~1).

While most existing research on fairness in policy learning focuses on value fairness, this paper extends a solution addressing action fairness. This paper is closely related to \citet{Frauen:2024}, who propose adjusting decision-relevant features to be independent of sensitive attributes using fair representation learning \citep{Zemel:2013}, and then learning a policy using a doubly robust score to guarantee action fairness. They also provide an option to promote value fairness by modifying the learning objective. A similar goal of achieving action fairness through variable transformation is pursued here, with the additional retention of variable interpretability to support transparency. This extension does not limit the potential to apply additional methods for ensuring value fairness among action fair assignments \citep[as proposed in][]{Frauen:2024}. As the lack of interpretability has been identified as a major barrier for the adoption of algorithmic decision making in practice, this approach addresses an important gap.

Further work on value fairness includes \citet{Tan:2022}, who adopt a so-called max-min fairness criterion by optimizing the mean of the lowest conditional outcomes (or another quantile) across sensitive groups, and \citet{Kock:2024}, who provide statistical guarantees for welfare maximization subject to the distributional equality of outcomes. Their framework supports diverse welfare functions, including quantiles and Gini-based metrics, enabling simultaneous targeting of fairness and distributional goals. Finally, \citet{Fang:2023} avoid the definition of a sensitive attribute and maximize average outcomes while requiring that a certain (user-specified) share of units benefits from the assigned treatment. A potential challenge for value fairness is the behavioral response of units receiving the assigned treatments, as realized outcomes depend on it. Incorporating potential non-compliance or strategic behavior into the decision-making algorithm in such settings might be necessary \citep{Shimao:2025}.

A complementary line of work by \citet{Viviano:2020} focuses on Pareto-efficient policies, i.e., those for which no alternative policy can improve outcomes for one group without worsening another. A fairness criterion then determines the least unfair choice among these.
In contrast, the procedure in this paper looks for a policy with the highest policy value under fairness constraints, rather than maximizing fairness under Pareto efficiency. This allows for scenarios where sacrificing policy value in one group may be acceptable to improve fairness. Unlike their approach, the sensitive attribute does not need to be binary.

\section{Conceptional Framework} \label{Setting}

\subsection{Notation}

The training data consists of $N$ i.i.d.~observations of the random vector $H = (D, Y, X)$, i.e.~$\{h_i=(d_i,y_i,x_i)\}_{i=1}^N$, drawn from an unknown probability distribution $\mathds{P}$. The observed outcome of interest is denoted by $Y$, with realizations $y_i \in \mathbb{R}$. The treatment variable $D$ is discrete, taking values $d_i \in \mathcal{D} = \{0, ..., M\}$.
The set of $G$ variables $X_{g}$ for $g \in \{1, \dots, G\}$ describes features of the observed units. They are collected in vector $X = (X_{1}, \dots, X_{G})$ with realizations $x_i \in \mathcal{X} \subseteq \mathbb{R}^G$.

To study fairness in a policy learning setting, some additional notation is needed. First, partition the vector of features into two groups of variables, $X = (S, Z)$. The first group, $S = (S_{1}, ..., S_{G_s})$, denotes a vector of features classified as sensitive by the decision maker. Information contained in these sensitive attributes must not have a direct effect on the treatment assignment. Formally, $S$ is required to be a discrete random vector with finite support, i.e.~with realizations $s_i \in \mathcal{S} \subset \mathbb{N}_0^{G_s}$, where $\mathcal{S}$ is of moderate cardinality. The remaining group consists of other features $Z = (Z_{1}, ..., Z_{G_z})$ with $G_z= G - G_s$. While all features $X$ are used to model the expected outcomes corresponding to each treatment, the treatment assignment rule 
may not be based on all of them. In particular, a vector of decision-relevant features $A$ is defined with realizations $a_i \in \mathcal{A} \subseteq \mathbb{R}^{G_a}$ and $G_{a} \leq G$.  The variables $A$ are used as inputs in the treatment assignment rule. Without fairness considerations, $A$ can be any subset of $X$.

\subsection{Policy Learning} \label{sec:optimal_policy_learning}
The framework considers off-policy learning from observational data with multiple treatments as described in \citet{Zhou:2022}. Let a policy be a decision rule $\pi: \mathcal{A} \mapsto \mathcal{D}$ mapping a unit's decision-relevant feature vector $a_i \in \mathcal{A}$ to a treatment $d_i \in \mathcal{D}$.
The expected reward from deploying a given policy $\pi$ is given by its policy value,  sometimes also referred to as welfare,
\begin{align*}
    V(\pi) = E\left[\Gamma_{\pi}(H)\right] \quad \text{with} \quad \Gamma_{\pi}(H) := \sum_{d \in \mathcal{D}} \mathbf{1}\{\pi(A)=d\} \Gamma_d(H)\text{,}
\end{align*}
where $\Gamma_d(H)$ denotes a score representing the value from assigning treatment $d$ to units with observed variables $H$. The goal of policy learning is to find the policy allocation $D^{\pi^*}$ which maximizes policy value, given a pre-specified policy class $\Pi$, i.e.,
\begin{align}\label{eq:optimal_policy}
    D^{\pi^*} = \pi^*(A) \quad \text{with} \quad \pi^* = \arg\max_{\pi \in \Pi} V(\pi)\text{.}
\end{align}
Restricting $\Pi$ is usually necessary to reduce complexity of the optimization problem \citep{Kitagawa:2018} or to ensure 
interpretability of the resulting policy. Examples of such policy classes include linear treatment assignment rules and finite-depth policy trees.

For $V(\pi)$ to have a causal interpretation, a suitable score and additional assumptions, such as randomization of treatments or unconfoundedness, are required. Under these conditions, with $Y^d$ denoting the potential outcome under treatment $d$ \citep{Rubin:1974}, the policy value can be expressed as
\begin{align} \label{eq:opt_pol_causal}
    V(\pi) = E\left[\sum_{d \in \mathcal{D}} \mathbf{1}\{\pi(A)=d\} Y^d\right]\text{.}
\end{align}
A common choice of a score that identifies \eqref{eq:opt_pol_causal} in causal settings is the individualized average potential outcome (IAPO),
\begin{align*}
    \Gamma_d^{\text{IAPO}}(H) = 
    \mu_d(X) \quad \text{with} \quad \mu_d(X) = \E[Y | D = d, X]\text{,}
\end{align*}
which represents the conditional mean outcome under treatment $d$. Another option is the augmented inverse probability-weighted (AIPW) score, or shortly doubly robust score, defined as
\begin{align*}
    \Gamma_d^{\text{AIPW}}(H) = \mu_d(X) + \frac{\mathbf{1}\{D=d\}(Y - \mu_d(X))}{e_d(X)} \quad \text{with} \quad e_d(x) = P(D=d| X = x)\text{,}
\end{align*}
which combines outcome modeling with inverse probability weighting. 

The AIPW score can have favorable asymptotic properties for policy learning, as demonstrated by \citet{Athey:2021} and \citet{Zhou:2022}. Let $\hat \pi$  be a learned policy solving $\hat \pi = \arg\max_{\pi \in \Pi} \frac{1}{N}\sum_{i=1}^N \sum_{d \in \mathcal{D}} \mathbf{1}\{\pi(a_i) =d\} \hat \Gamma_d(h_i)$ and $\hat \Gamma_d$ be estimated scores. Under suitable assumptions and an appropriate choice of the policy class, \citet{Zhou:2022} show that regret, defined as the difference between the value of the best policy and the learned policy $V(\pi^*) - V(\hat \pi)$, where $\hat \pi$ is based on estimated AIPW scores, attains the asymptotically minimax-optimal rate. This result applies to a range of policy classes, including finite-depth policy trees introduced in Section~\ref{sec:pol_trees}, under bounded outcomes and mild conditions on nuisance parameter estimators based on cross-fitting.

Nonetheless, the action fairness procedures studied in this paper operate on general scores that are functions of $H$. This flexibility is possible by the structure of the proposed approach, in which fairness constraints are imposed after the estimation of the scores but prior to the policy learning step, by transforming the decision-relevant features. Therefore, the specific choice of the score is not consequential for targeting fairness. Note that without causal assumptions, $V(\pi)$ represents a predictive score-based policy value, which may differ from the true causal policy value defined in terms of potential outcomes.

\subsection{Policy Trees}
\label{sec:pol_trees}
Policy trees, introduced by \citet{Athey:2021} and \citet{Zhou:2022}, form a class of policy functions based on decision trees. Like traditional decision trees \citep{Breiman:1984}, policy trees assign treatments by partitioning units into subgroups based on their feature values, following paths from the root node down to the leaf nodes. However, rather than minimizing predictive errors based on observed outcomes, policy trees explicitly aim to maximize the policy value, as formulated in equation (\ref{eq:optimal_policy}).

Several features make policy trees an appealing choice for policy learning. First, they balance interpretability and policy value maximization. Their hierarchical structure enables decision makers to transparently follow the decision logic, significantly enhancing interpretability of the model. Additionally, unlike traditional decision trees, which rely on greedy algorithms and locally optimal splits, policy trees employ exact optimization. This exhaustive search for the globally optimal tree structure aims to yield a policy that maximizes the policy value, albeit at higher computational cost.

Motivated by these desirable properties, the methodology of policy trees is adapted to the fairness-aware setting in Section \ref{sec:pre_poc}. In terms of equation (\ref{eq:optimal_policy}), this means restricting the policy class $\Pi$ to finite-depth policy trees. While the exposition focuses on policy trees, the general principles extend to other threshold-based policy classes. 

\subsection{Interpretability}
While policy trees offer a clear decision structure, interpretability itself is a broader concept. While there is no universal mathematical definition of interpretability, it generally refers to the \textit{``degree to which a human can understand the cause of a decision''} \citep{Miller:2019, Molnar:2020}. A straightforward way to promote interpretability is by using so-called interpretable decision-making models, such as policy trees or other threshold-based rules, where variables are compared against numerical thresholds to assign treatments.

Yet, model structure alone does not guarantee interpretability. If the decision-relevant variables lack intuitive meaning, e.g.~due to transformation, the thresholds may not be meaningful to human decision makers. Therefore, we define interpretability as consisting of two complementary components: \textit{model interpretability}, which refers to the intrinsic interpretability of the model structure, and \textit{variable interpretability}, which refers to the understandability of the variables used in the decision-making process.

\section{Pre-Processing for Policy Learning with Sensitive Attributes}\label{sec:pre_poc}

As mentioned, this study extends a pre-processing approach to promote action fairness in policy learning while ensuring interpretability. Instead of altering the optimization algorithm itself, fairness is targeted by pre-processing the input data, allowing the optimization routine to remain unchanged. As the pre-processing step may compromise variable interpretability, the parameters of the resulting policy rule are transformed to the original scale of the decision-relevant features to restore it. Before explaining each step in detail, an overview of the approach is presented.

\subsection{The Policy Learning Pipeline}\label{sec:pipeline}

Figure \ref{fig:pipeline_fair_policy_learning} illustrates the proposed procedure. Given an i.i.d.~sample $\{h_i\}_{i=1}^N$, a score $\Gamma_d(H)$ is estimated for each treatment $d \in \mathcal{D}$. The features $X$ include all features required for the score estimation, including sensitive attributes. In the second step, the decision-relevant features $A$ and/or the estimated scores $\hat \Gamma_d(H)$ are transformed to fairness-adjusted versions, $\tilde A$ and $\tilde \Gamma_d(H)$, using the pre-processing procedure described in Section~\ref{attaining_action_fariness}. Using the transformed data, a policy is learned by applying a standard interpretable policy learning algorithm. While this ensures model interpretability, the transformation of variables may compromise variable interpretability. To restore variable interpretability, the parameters of the learned policy are mapped back to the scale of the original decision-relevant features separately for each sensitive group (see Section~\ref{keeping_explainability}). The final policy function can then be applied to a new observation $a_i$ to yield a recommended treatment $d_i^{\hat \pi}$. As a result of the pre-processing step and the transformation of the policy parameters, the policy recommendations are fully interpretable and promote action fairness (see Definition~\ref{def:action_fairness} in Section~\ref{sec:fair_def}). The essential component is the pre-processing step that modifies the data to target fairness. Without it, the procedure reduces to a fairness-unaware policy learning framework, represented by dashed lines in Figure \ref{fig:pipeline_fair_policy_learning}.

\begin{figure}[htbp!]
\caption{Illustration of the pipeline for interpretable policy learning with sensitive attributes}\label{fig:pipeline_fair_policy_learning}
    \begin{center}
\begin{tikzpicture}[
block/.style={
draw,
fill=white,
rectangle, 
minimum width=2.5cm,
minimum height=1.5cm,
font=\footnotesize,
align=center}]
\node[block](data){\textbf{Data}\\$\{h_i\}_{i=1}^N$};
\node[block,right=1.5cm of data](score){\textbf{Score}\\$\hat \Gamma_d(H)$};
\node[block,below=1.5cm of score](adjscore){{\textbf{Adjusted}}\\{\textbf{data}\par}\\ {$\tilde A$, $\tilde \Gamma_d(H)$}};
\node[block,right=1.5cm of adjscore](policy){\textbf{Fairness-} \\ {\textbf{aware policy}}\\$ {\hat \pi}(\tilde A)$};
\node[block,right=1.5cm of score](expolicy){\textbf{Interpretable}\\\textbf{policy}\\$ {\hat \pi}(A)$};
\node[block,right=1.5cm of expolicy](decision){\textbf{Treatment}\\$ d_i^{\hat \pi}$};
\node[block,right=1.5cm of policy](newobs){\textbf{New}\\\textbf{observation}\\$a_i$};
\draw [arrow, dashed] (data) -- (score) node[midway, above, font=\scriptsize, align=center] {estimate};
\draw [arrow] ([yshift=-.3cm]data.east) -- ([yshift=-.3cm]score.west) node[midway, above, font=\scriptsize, align=center] {};
\draw [arrow] (score) -- (adjscore) node[midway, left, font=\scriptsize, align=right] {pre-\\process};
\draw [arrow] (adjscore) -- (policy) node[midway,above] {\scriptsize optimize};
\draw [arrow] (data) -- (adjscore)  node[midway, left, font=\scriptsize, align=center] {pre-\\process};
\draw [arrow] (policy) -- (expolicy) node[midway, right, font=\scriptsize, align=left] {trans-\\form};
\draw [arrow, dashed] (expolicy) -- (decision) node[midway,above] {\scriptsize predict};
\draw [arrow] ([yshift=-.3cm]expolicy.east) -- ([yshift=-.3cm]decision.west) node[midway, above, font=\scriptsize, align=center] {} node[midway,below] {\scriptsize for $a_i$};
\draw [arrow, dashed] (score) -- (expolicy) node[midway,above] {\scriptsize optimize};
\draw [arrow, dashed] ([xshift=.15cm]newobs.north) -- ([xshift=.15cm]decision.south) node[midway, right, font=\scriptsize, align=left] {predict};
\draw  [arrow] ([xshift=-.15cm]newobs.north) -- ([xshift=-.15cm]decision.south);
\end{tikzpicture}
\end{center}
\vspace{0.2cm}
\caption*{\scriptsize \textit{Notes:} A score $\Gamma_d(H)$ is estimated from data $\{h_i\}_{i=1}^N$, adjusted for fairness, and used alongside adjusted decision-relevant features $\tilde A$ to optimize the fairness-aware policy $\hat \pi(\tilde A)$. The fairness-aware policy is transformed to recover the variable interpretability and evaluated at $a_i$ to derive recommended treatment $d_i^{\hat \pi}$. Solid lines represent the proposed fairness-aware framework, dashed lines represent fairness-unaware framework without fairness-adjustment. Note that the transformation of the fairness-aware policy function $\hat \pi(\tilde A)$ leads to a fairness-aware interpretable policy $\hat \pi(A)$ that will include $S$ into $A$ as described in Section \ref{keeping_explainability}.} 
\end{figure}

\subsection{Definition of Fairness}\label{sec:fair_def}
This paper follows the definition of action fairness from \citet{Frauen:2024}, which adapts the statistical parity criterion, a standard fairness notion in machine learning \citep{Barocas:2017}:

\begin{definition}\label{def:action_fairness} \text{(Action Fairness)}
A policy $\pi \in \Pi$ is action fair if the treatment assignment $D^\pi$ generated by $\pi$ is independent of the sensitive attributes, i.e.~$D^\pi \ind S$ for $D^{\pi} = \pi(A)$.
\end{definition}

It requires that treatments recommended by the policy are independent of sensitive attributes, thereby promoting equal opportunity by providing all sensitive groups with the same chance of receiving treatment.\footnote{Definition \ref{def:action_fairness} implies that $S$ cannot influence recommended treatments 
$D^{\pi}$. A less restrictive variant requires independence only conditional on so-called materially relevant variables \citep{Strack:2023}. This would allow $S$ to influence $D^{\pi}$ indirectly through these variables, analogous to conditional statistical parity. 
Extending the proposed methods to this broader fairness criterion is beyond the scope of this paper.}
This fairness concept is particularly relevant when the direction of the treatment effect has the same sign across groups, even if the magnitude differs.

To illustrate treatment assignment under action fairness, consider an example where the objective is to allocate unemployed individuals to vocational training programs and language courses. The sensitive attribute is citizenship, categorized as either ``domestic'' or ``foreign''. Assume that foreigners are already highly qualified, such that natives benefit significantly more from vocational training. Meanwhile, foreigners benefit more from language courses than natives, as improved language skills facilitate their entry into the labor market. In this context, an unrestricted fairness-unaware policy would assign all natives to the vocational training program and all foreigners to language courses, as this would maximize the policy value. However, a policy based on action fairness would allocate similar proportions of vocational training and language course slots to both natives and foreigners, taking fairness in the access to the programs into account when maximizing policy value.

Action fairness may not be a meaningful concept in every context. For instance, \citet{Dwork:2012} argue that statistical parity fails in classification settings where outcomes are reversed between sensitive groups. In policy learning, similar problems could occur, for example, if a certain program type is targeted towards the needs of a particular sensitive group. To illustrate, consider language courses that exclusively teach the local language. For native speakers, such assignments may have adverse effects, as they are placed in a program that does not meet their needs, prolonging their period of unemployment. Assigning equal shares of natives and foreigners does not appear to be a reasonable choice in this case. Therefore, it is crucial that the selection of the sensitive attributes and the fairness criterion is aligned with the specific task at hand.

\subsection{Attaining Action Fairness by Pre-Processing} \label{attaining_action_fariness}

To develop pre-processing procedures, it is first necessary to determine which inputs to the policy learning algorithm should be transformed when targeting action fairness. In this setting, the decision-relevant features $A$ are taken as a subset of $Z$, and the transformation relies on the following relationship:

\begin{lemma}(Independence of decision-relevant features implies independence of assigned treatments)\\
    Let $\tilde A$ be a random variable such that $\tilde A \ind S$. Then it holds that $D^{\pi} \ind S$ for $D^{\pi} = \pi(\tilde A)$, where $\pi$ can be any policy which is a function of $\tilde A$ only \citep[e.g.][Theorem 4.3.5]{casella2002statistical}.
\end{lemma}

Hence, if (adjusted) decision-relevant features $\tilde A$ are jointly independent of the sensitive attributes $S$, then treatment assignments $D^{\pi}$ will be independent of $S$, when the policy is a function of $\tilde A$. One straightforward pre-processing approach would then select decision-relevant features $\tilde A$ as the subset of features from $Z$ that are jointly independent of $S$. 
This could be tested before running the policy learning algorithm by performing independence tests. However, such an approach would be restrictive in real world applications, as many variables often correlate with each other. A more flexible approach is to transform the observed decision-relevant features $A$ to a version $\tilde A$ that is jointly independent of $S$ while keeping as much information from the original $A$ as possible. This is one of the principal ideas of fairness pre-processing in the algorithmic fairness literature \citep[e.g.][]{Johndrow:2019, Feldman:2015}. 
Formally, the objective is to select $\tilde A$ from the set of all transformations satisfying the independence condition $\tilde A  \ind S$, such that the distance to the original decision-relevant features, $\Delta(A, \tilde A)$, is minimized.\footnote{This objective can be viewed through the lens of an optimal transport problem, which looks for the most efficient way to transform one probability distribution into another by minimizing a cost of moving “mass.” The Wasserstein distance, introduced in \citet{vaserstein:1969}, quantifies this minimal cost when the transport cost is a power of the Euclidean distance, and thus plays a key role in determining optimal couplings between distributions. In the univariate case, this corresponds to matching quantiles.}
Once the transformed $\tilde A$ is obtained, policy learning can proceed in the usual way.

To introduce the procedure for obtaining the adjusted features $\tilde A$, consider first the case where $A$ is a univariate continuous variable. Let $F_{A|S}(A,S)$ denote the conditional cumulative distribution function (cdf) of $A$ given the sensitive attributes $S$. \citet{Johndrow:2019} suggest the variable transformation $\tilde A = F_A^{-1}(F_{A|S}(A,S))$, where $F^{-1}_{A}(\cdot)$ denotes the (marginal) quantile function of $A$, to produce a variable $\tilde A$ that is independent of $S$ (as formalized in Lemma~\ref{quant_lemma}). They also show that this transformation coincides with the optimal transport map solving the corresponding one-dimensional optimal transport problem with a Euclidean cost.\footnote{Regarding further optimality results in the literature, \cite{Strack:2023} show a few special cases where $\tilde A$ is the best \textit{privacy-preserving signal} of $A$ for utility maximizing decision problems.} 
The variable $\tilde A$ can thus be interpreted as a fairness-adjusted version of $A$ that preserves its marginal distribution across sensitive groups while removing statistical dependence on $S$. In the remainder of the paper, the procedure based on first applying the conditional cdf $F_{A|S}(\cdot,\cdot)$ and then the marginal quantile (MQ) function $F^{-1}_{A}(\cdot)$ to obtain $\tilde A$ is referred to as \emph{MQ-adjustment}.

\begin{lemma} \label{quant_lemma}
Let $A$ be a univariate continuous random variable. The adjusted variable, defined as $ \tilde A = F^{-1}_{A}\left( F_{A|S}(A, S) \right)$ is statistically independent of $S$.\\
\textbf{Proof:} The conditional cdf $F_{A|S}(A, S)$ returns a random variable uniformly distributed on the interval $[0, 1]$  that is independent of $S$. This property is preserved when applying the quantile function $F^{-1}_{A}$ as this transformation does not depend on $S$.
\end{lemma}

In practice, the conditional cdf $F_{A|S}(A, S)$ is unknown. Since $S$ is required to be discrete with low cardinality, a nonparametric estimate of $F_{A|S}(A, S)$ can be obtained by computing the empirical cdf of each observation $a_i$ within the sensitive group $s$. There are various definitions of the empirical cdf. In this paper, Definition~7 in \citet{Hyndman:1996} is followed. This version maps the sample minimum to $0$, the maximum to $1$, and assigns evenly spaced probabilities to the remaining order statistics. The same approach is used to estimate the marginal cdf $F_A(\cdot)$. Probabilities from the estimated $F_{A|S}(\cdot,\cdot)$ are then transformed to $\tilde A$ using the estimated marginal quantile function, $F^{-1}_A(\cdot)$. This procedure enforces that the empirical conditional cdfs $F_{\tilde A|S}(\cdot,\cdot)$ are identical across sensitive groups, i.e., $\tilde A$ is (empirically) independent of $S$. At the same time, within-group ranks and the empirical marginal distribution of $A$ are preserved.

For discrete random variables or any random variables with mass points, the adjustment is slightly more involved. In such cases, the approach of \citet{Johndrow:2019} is followed, which assigns random values from the corresponding conditional cdf intervals to observations with tied values.\footnote{Note that discrete decision-relevant features do not need to have inherent ordering. For ease of exposition, the rest of the paper focuses on ordered categorical variables.} In particular, let $\dot{ \mathcal{A}} = \{ \dot{a}_1, \dot{a}_2, ...\}$ be the set of all values that $A$ can take, listed in increasing order such that $\dot{a}_{m-1} < \dot{a}_m$. The adjusted feature is then defined as 
\begin{align*}
    \tilde A = F_{A}^{-1}(\zeta(A, S)) \quad \text{with} \quad \zeta(A, S) | A = \dot{a}_m \sim \text{Uniform}\left(F_{A|S}(\dot{a}_{m-1},S), F_{A|S}(\dot{a}_{m},S)\right)\text{,}
\end{align*}
where $\dot{a}_0 = -\infty$. \citet{Johndrow:2019} show that this transformation has the same optimal transport property as the continuous case. Algorithm~\ref{alg:adjust} summarizes the procedure for handling both continuous and discrete decision-relevant features, building on Algorithm~1 in \citet{Johndrow:2019} and adapting it to the specific requirements of our framework.

\begin{algorithm}[ht]
\caption{\textsc{MQ-adjustment}}\label{alg:adjust}
\KwIn{Univariate decision-relevant variable $\{a_i\}_{i=1}^N$ and sensitive attributes $\{s_i\}_{i=1}^N$}
\KwOut{Values of the empirical cdf $\{p_i\}_{i=1}^N$ and fairness-adjusted decision-relevant variables $\{\tilde{a}_i\}_{i=1}^N$}
\For{$s \in \mathcal{S}$}{
    Let $\mathcal{I}_s = \{ i : s_i = s \}$\;
    \For{$i \in \mathcal{I}_s$}{
        Compute $Rank(a_i) = \sum_{j \in \mathcal{I}_s} \mathbf{1}\{a_j \leq a_i\}$\;
        Compute $p_i = \frac{Rank(a_i)-1}{|\mathcal{I}_s|-1}$\;
        \If{$| \{ k \in \mathcal{I}_s : a_k = a_i \} | > 1$}{
            $\mathcal{M}_{intersec} = \{m: \dot{a}_m \in \dot{\mathcal{A}} \cap \mathcal{A}_s  \}$, where $\mathcal{A}_s = \{a_i: s_i =s\}$\;
            Set $a_i^{lower} = \max\{\dot{a}_m: \dot{a}_m < a_i, m \in \mathcal{M}_{intersec}\}$\;
            \If{$a_i^{lower}=\emptyset$}{$a_i^{lower}=-\infty$\;}
            Compute $p^{lower}_i = \frac{1}{|\mathcal{I}_s|-1} \sum_{j \in \mathcal{I}_s} \mathbf{1}\{a_j \leq a_i^{lower}\}$\;
            Update $p_i = \text{Uniform}(p^{lower}_i, p_i)$\;
        }
    }
}
Let $\{a_{(1)}, \dots, a_{(N)}\}$ be all $\{a_i\}_{i=1}^N$ sorted in ascending order\;
\For{$i = 1, ..., N$}{
    Let $c_i= 1 + (N-1)p_i$\;
    Set $\lambda=\lfloor c_i\rfloor$ and $\kappa = c_i-\lambda$\;
    Compute empirical quantile $\tilde a_i = (1-\kappa)a_{(\lambda)} + \kappa a_{(\lambda+1)}$\;}
\end{algorithm}

The \emph{MQ-adjustment} seems attractive as it achieves statistical independence between a decision-relevant feature and the sensitive attributes, while remaining as close as possible to $A$ in distribution. However, when applied to multiple decision-relevant variables individually, this procedure only achieves pairwise independence and not joint independence, i.e.~for two decision-relevant features $A_{1}$ and $A_{2}$, the result is $\tilde A_{1} \ind S$ and $\tilde A_{2} \ind S$ but not necessarily $\{\tilde A_{1}, \tilde A_{2} \} \ind S$. To solve this problem, \citet{Johndrow:2019} propose chained adjustments based on the decomposition 
\begin{align*}
    F_{A_{1}, A_{2}|S}(A_{1}, A_{2}, S)=F_{A_{1}|A_{2}, S}(A_{1}, A_{2}, S)F_{A_{2}|S}(A_{2}, S)\text{.}
\end{align*}
Then, the cdfs $F_{A_{2}|S}(A_{2}, S)$ and $F_{A_{1}|A_{2}, S}(A_{1}, A_{2}, S)$ are used to adjust the decision-relevant features instead of $F_{A_{1}|S}(A_{1}, S)$ and $F_{A_{2}|S}(A_{2}, S)$. However, these distributions are very hard to estimate in practice without relying on strong assumptions, especially if continuous variables appear in the conditioning sets. Hence, in the application below, the \emph{MQ-adjustment} is applied separately to each decision-relevant feature, at the cost of not attaining joint independence.

\subsection{Preserving Interpretability in Policy Trees} \label{keeping_explainability}

The \emph{MQ-adjustment} leads to the loss of variable interpretability when the adjusted features are used in interpretable policy learning algorithms such as policy trees.
To address this, a transformation back to the original scale is proposed using the conditional quantile (CQ) function $F_{A|S}^{-1}(\cdot, \cdot)$. This transformation not only maps values back to the scale of $A$ but also recovers its value exactly, since it is the inverse of the transformation used to obtain $F_{A|S}(A, S)$ and consequently $\tilde A$. 
Applying this transformation to splitting thresholds expressed on the cdf-scale is proposed to restore the interpretability of policy trees and is referred to as \emph{CQ-adjustment}.

Note that both $\tilde A$ and $F_{A|S}(A,S)$ (including the randomized values $\zeta(A,S)$ for discrete $A$) can be used to build a fairness-aware policy tree. The cdf-scale threshold $p$ supplied to the CQ-function may be either (i) a splitting value from a tree optimized directly on $F_{A|S}(A,S)$, or (ii) a transformed value $p = F_A(\tilde a)$ from a tree optimized on $\tilde A$. For ease of implementation, the first option is adopted in the application, i.e., when building a policy tree, splits are performed directly on $F_{A|S}(A,S)$, which avoids applying $F_A^{-1}(\cdot)$ in the \emph{MQ-adjustment} and then converting thresholds $\tilde a$ back to the corresponding cdf values $p = F_A(\tilde a)$ for the \emph{CQ-adjustment}.\footnote{For continuous features $A$, splitting on $\tilde A$ or $F_{A|S}(A, S)$ is equivalent. For discrete features $A$, splitting on randomized values $\zeta(A,S)$ instead of $\tilde A$ provides additional flexibility by offering a larger set of potential splitting points.} 
To obtain splitting thresholds on the original scale, \emph{CQ-adjustment} applies the transformation $g(p, s) = F^{-1}_{A|S}\left(p, s\right)$ in each node for all $s \in \mathcal{S}$. As a result, the original fairness-aware policy tree is translated to separate trees for each sensitive group, which are fully interpretable in terms of the original features $A$. To obtain the interpretable fairness-aware assignment for a given unit, the appropriate group-specific tree is selected based on the unit's sensitive attributes and then evaluated using the observed decision-relevant features.\footnote{Alternatively, the group-specific trees can be condensed into a single tree that first splits on sensitive attributes and then continues with the group-specific policy trees (see Figure \ref{fig:probtree_extleft_depth_3}).}

The implementation of \emph{CQ-adjustment} involves two key challenges. The first concerns the practical computation of the transformation $F^{-1}_{A|S}(p, s)$. Given a cdf-scale threshold $p$, a sensitive group $s$, the original decision-relevant features $\{a_i\}_{i \in \mathcal{I}_s}$ in the group $s$ and their empirical conditional cdf values $p_i = F_{A|S}(a_i, s)$, the algorithm proceeds as follows. If $p$ matches one of the values $\{p_i\}_{i \in \mathcal{I}_s}$, the corresponding $a_i$ is taken directly as the transformed threshold. Otherwise, the nearest $p_i$ values below and above $p$ are found, along with their associated $a_i$ values, and linear interpolation is performed. Specifically, the relative position of $p$ between the two $p_i$ values is computed and applied to interpolate between the corresponding $a_i$ values. The complete procedure is summarized in Algorithm~\ref{alg:lookup}. For continuous decision-relevant features, this method guarantees that the resulting group-specific policy trees yield the same treatment assignments as those based on the empirical cdf values $p_i$.

\begin{algorithm}[ht]
\caption{\textsc{CQ-adjustment}}\label{alg:lookup}
\KwIn{Cdf-scale threshold $p$, sensitive group $s$ and pairs $\{(a_i, p_i)\}_{i \in \mathcal{I}_s}$, 
  where $a_i$ are observed values of a univariate decision-relevant variable $A$ in group $s$, $p_i$ their empirical conditional cdf values $F_{A|S}$ and $\mathcal{I}_s = \{ i : s_i = s \}$}
\KwOut{Transformed threshold on original scale $g(p,s)$}
\eIf{$p \in \{ p_i : i \in \mathcal{I}_s \}$}{
Choose the $k \in \mathcal{I}_s$ such that $p_k = p$\;
Set $g(p,s) = a_k$\;
}{
Let $i_{\text{lower}} = \arg\max\{ p_i : p_i < p,\, i \in \mathcal{I}_s \}$\;
Let $i_{\text{upper}} = \arg\min\{ p_i : p_i > p,\, i \in \mathcal{I}_s \}$\;
Set $p_{\text{lower}} = p_{i_{\text{lower}}}$ and $p_{\text{upper}} = p_{i_{\text{upper}}}$\;
Set $a_{\text{lower}} = a_{i_{\text{lower}}}$ and $a_{\text{upper}} = a_{i_{\text{upper}}}$\;
Compute $g(p,s) = a_{\text{lower}} + \left( \frac{a_{\text{upper}}-a_{\text{lower}}}{p_{\text{upper}}-p_{\text{lower}}}\right) (p - p_{\text{lower}})$\;
}
\end{algorithm}

To illustrate, consider a case where the splitting threshold is $p=0.33$ and the objective is to transform this value for the sensitive group $s=1$. Assume there exists an individual $i=13$ with $s_{13}=1$, $p_{13}=0.33$ and $a_{13}=2$. Then, the value of $a_i$ for this individual can be used directly to transform the threshold, yielding $g(p, s) = a_{13} = 2$. Now consider the case where no individual with $s_i = 1$ has a value of $p_i$ exactly equal to $0.33$. In this situation, two observations in the group are identified, one just below and one just above the threshold $p$, for example, $p_{29} = 0.32$ and $p_{17} = 0.35$, with corresponding original values $a_{29} = 1.5$ and $a_{17} = 3$. Then a linear interpolation is applied to compute the transformed threshold as $g(p,s) = 1.5 + \left(\frac{3-1.5}{0.35-0.32}\right)(0.33-0.32) = 2$.

The second challenge arises when some decision-relevant features are discrete. In this case, units with the same value of $A$ are assigned  randomized values from their empirical cdf interval, as specified in the \emph{MQ-adjustment}.  Therefore, when transforming a cdf-scale threshold $p$ back to the original scale, a range of empirical cdf values maps to a single value $a$.  As a result, the proportion of observations falling on either side of the split based on threshold $a$ may differ from that in the tree based on threshold $p$. To address this issue, the concept of a probabilistic split is introduced. For a given  cdf-scale splitting threshold $p$ and sensitive group $s$, the threshold  is first mapped to the original scale using \emph{CQ-adjustment} (Algorithm~\ref{alg:lookup}). Next,  among the observations in the current leaf with $a_i = g(p,s)$ and $s_i = s$, the share that would have been assigned to each child node in the tree based on $p$ is computed. These shares represent the probabilistic split information and are used for evaluations of the tree.

To continue the previous example, suppose the threshold $p=0.33$ maps to the original value $g(p,1)=2$ for individuals in the sensitive group $s=1$. Assume that within the evaluated leaf, ten observations share the combination $a_i=2$ and $s_i=1$. Of these, eight have $p_i<0.33$ and two have $p_i>0.33$. To ensure that the partition based on $a$ matches the split implied by $p$, all observations with $a_i<2$ and 80\% of those with $a_i=2$ are assigned to the left child node, while the remaining 20\% with $a_i=2$, along with all observations with $a_i>2$, proceed to the right child node. Implementing this rule in practice requires a random draw to determine the branch for individuals exactly at the threshold, which makes the split probabilistic. Due to the inherent randomness in these splits, the resulting assignments may slightly differ each time the tree is evaluated. Pseudo-code of the procedure is presented in Algorithm \ref{alg:probabilistic_tree}, while Figure~\ref{fig:probtree_extleft_depth_3} offers a visual illustration of a probabilistic split tree as part of the empirical application.

\begin{algorithm}[ht]
\caption{\textsc{ProbSplitTree}}\label{alg:probabilistic_tree}
\KwIn{Policy tree $\text{Tree}(p)$ based on empirical conditional cdf, observations $\{(a_{ij}, p_{ij}, s_i): i=1,...,N, j=1,...,G_a\}$}
\KwOut{Transformed group-specific trees $\text{Tree}(a|s)$}
\For{$s \in \mathcal{S}$}{
    Initialize $\text{Tree}(a|s) = \text{Tree}(p)$\;
    \For{node $l$ in $\text{Tree}(p)$}{
        Let $\mathcal{L}$ be the set of indices $i$ in node $l$\;
        Let $j$ be the index of the decision-relevant feature used for splitting in node $l$\;
        Let $p$ be the cdf-scale splitting threshold in node $l$\;
        Let $\mathcal{I}_s = \{ i : s_i = s \}$\;
        Apply Algorithm \ref{alg:lookup}: $g(p,s) = \textsc{CQ-adjustment}\left(p, s, \{(a_{ij}, p_{ij}): i \in \mathcal{I}_s\}\right)$\;
        Let $\mathcal{I} = \{ i \in \mathcal{L} : a_{ij} = g(p,s), s_i = s\}$\;
        Compute $\tilde p = \frac{1}{|\mathcal{I}|} \sum_{i \in \mathcal{I}} \mathbf{1}(p_{ij} \leq p$)\;
        \eIf{$\tilde p = 1$}{
            Update split condition at node $l$:\\
            \Indp $A_{j} \leq g(p,s)$ for continuous $A$\;
            $A_{j} \leq \lfloor g(p,s) \rfloor$ for discrete $A$\;
        }{
            Update split condition at node $l$: \\
            \Indp $A_{j} < g(p,s)$, and \\
            $\tilde p$ share of units with $A_{j} = g(p,s)$\;
            \Indm
        }
    }
}
\end{algorithm}

The following two lemmas formalize the two variants of the \emph{CQ-adjustment}. Lemma \ref{lem:unique_split} shows that applying $F^{-1}_{A|S}(p,s)$ to yield splitting thresholds in terms of $A$ is well-defined and reverses the \emph{MQ-adjustment} in the continuous case. Lemma \ref{lem:prob_split} shows that \emph{CQ-adjustment} extended by a concept of probabilistic split is well-defined and reverses \emph{MQ-adjustment} in the discrete case.

\begin{lemma}\label{lem:unique_split}
Let $S$ be a discrete random variable and $F_{A|S}(a,s)$ be a continuous and strictly monotone conditional cdf in $a$ for all $s \in \mathcal{S}$. Let $\tilde A$ be obtained via MQ-adjustment, i.e., $\tilde A = F_A^{-1}(F_{A|S}(A,S))$. Then for every $\tilde a$ and given $s$, with $F_A(\tilde a) = p$, there exists a unique mapping $a = g(p,s)$ such that $F_{A|S}(a,s) = p$.
\\
\textbf{Proof}: By construction of the MQ-adjustment, $F_A(\tilde a) = F_{A|S}(a,s) = p$. 
The strict monotonicity and continuity of $F_{A|S}(\cdot,s)$ ensure that $a$ is uniquely determined by $a = g(p,s) = F^{-1}_{A|S}(p,s)$. 
\end{lemma}

\begin{lemma}\label{lem:prob_split}
Let $S$ be a discrete random variable and, for each $s\in\mathcal S$, let $A\mid S=s$ have finite support $\dot{\mathcal A}=\{\dot a_1<\dot a_2<\dots\}$ with conditional cdf $F_{A\mid S}(\cdot,s)$. Let $\tilde A$ be obtained via MQ-adjustment, i.e., $ \tilde A = F_{A}^{-1}(\zeta(A, S)) \text{ with } \zeta(A, S) | A = \dot{a}_m \sim \text{Uniform}\left(F_{A|S}(\dot{a}_{m-1},S), F_{A|S}(\dot{a}_{m},S)\right)$. Fix a leaf $l$ and group $s$. Then for any $\tilde a$ with $F_A(\tilde a) = p$, there exists a unique index $m$ such that $ F_{A|S}(\dot a_{m-1},s) < p \le F_{A|S}(\dot a_m,s)$, and a unique reverse mapping $(a, \tilde p) = g(p,s)$ with $a=\dot a_m$ and $\tilde p$ being uniquely determined in terms of leaf shares $$\tilde p = Pr\big(\zeta(A, s)\le p \,\big|\, A=\dot a_m,\; S=s,\; (A,S)\in l\big),$$
which ensures a well-defined reverse of the corresponding MQ-adjustment. \\

\textbf{Proof:}
Right-continuity and monotonicity of $F_{A|S}(\cdot,s)$ imply that, for any $p = F_A(\tilde a)$ and fixed $s$, there exists a unique index $m$ with $F_{A|S}(\dot a_{m-1},s) < p \le F_{A|S}(\dot a_m,s)$. Hence the corresponding threshold on the original scale is $a = \dot a_m$. To recover the split based on $\tilde a$ (or $p$), the reverse mapping needs to determine the fraction of leaf observations with $a_i = \dot a_m$ and $s_i =s$ whose $\zeta$-values fall below $p$. This share is $\tilde p = Pr\big(\zeta(A, s)\le p \,\big|\, A=\dot a_m,\; S=s,\; (A,S)\in l\big)$, which uniquely determines the probabilistic split at $a = \dot a_m$ in leaf $l$ due to the uniform distribution of $\zeta(A,s)$.
\end{lemma}

\subsection{Implementing Fairness-Aware Policy Learning in Practice} \label{Heuristics}

As pointed out in the previous subsection, the proposed procedure achieves action fairness for a univariate decision-relevant feature. However, such a setting is rarely  encountered in practice,  where multiple decision-relevant features of different types  are typically involved. As discussed in Section \ref{attaining_action_fariness},  one possible approach is to impose simplifying assumptions and parametric modeling to achieve joint independence in such cases. Alternatively, this paper proposes to  relax the objective of achieving joint independence. Instead, the researcher or decision maker can use one of the following three heuristics to approximate action fairness in empirical applications while retaining interpretability of policy trees. Since independence of the resulting assignments can be tested, the decision maker can select the procedure achieving the best balance between action fairness and maximization of the policy value in a particular application.

\textbf{Pairwise MQ-CQ-adjustment of decision-relevant features:} If several decision-relevant features are present, each can be adjusted individually using \emph{MQ-adjustment}.  Although this does not guarantee joint independence from sensitive attributes, fairness is expected to improve as the marginal distribution of each decision-relevant feature is rendered independent of $S$. Remaining dependencies may still transmit some sensitive information, yet the overall influence of sensitive attributes on the final decision shall be reduced. This idea is already put forward in \citet{Feldman:2015} for the classification setting. Interpretability can be retained by using the \emph{CQ-adjustment} (in combination with the described probabilistic split tree for discrete features).

\textbf{Pairwise MQ-adjustment of scores:} Another possible approach would be to adjust the scores instead of the decision-relevant features.
Specifically, scores are pre-processed using \emph{MQ-adjustment} yielding $[\tilde \Gamma_0, ..., \tilde \Gamma_M]$ while leaving $A$ unadjusted. The optimization problem can then be solved using $[\tilde \Gamma_0, ..., \tilde \Gamma_M]$ and $A$. This procedure may improve fairness by making the scores pairwise independent of sensitive attributes. However, fairness improvement is not guaranteed because the resulting decisions remain functions of $A$. Therefore, empirical tests of independence are recommended to evaluate the fairness of the treatment assignments. The advantage of this procedure is that the policy tree remains interpretable without adjustments since it is optimized using the original decision-relevant features $A$.

\textbf{Pairwise MQ-CQ-adjustment of decision-relevant features and pairwise MQ-adjust\-ment of scores:} Finally, the two previously described heuristics can be combined.

\section{Empirical Application}\label{Empirical_example}

To showcase the proposed methods in an empirical application, algorithmic assignments of unemployed individuals to Active Labor Market Policies (ALMP) in Switzerland are analyzed. There has been growing interest in enhancing allocation of unemployed individuals to labor market programs. An early study using Swiss data by \citet{Lechner:2007} demonstrates that statistical treatment rules may outperform caseworker decisions in assigning individuals to programs. In more recent work, \citet{Knaus:2022b} and \cite{Cockx:2023} apply policy learning to improve program targeting, following the framework established by \citet{Zhou:2022}. \cite{Cockx:2023} find that allocations based on shallow policy trees yield better outcomes than both observed caseworker assignments and random allocation in the Belgian setting. Building on this line of research, \cite{Mascolo:2024} utilize causal machine learning with Swiss administrative data to assess the medium-term effects of ALMP on employment and earnings.
Their results further reinforce the value of shallow policy trees in optimizing program assignment. While fairness considerations have previously been used to justify the choice of decision-relevant features in policy trees \citep[e.g.][]{Knaus:2022b}, research systematically addressing fairness in  assignment of unemployed to ALMP remains limited. Recent contributions based on Swiss administrative data include \citet{zezulka:2024}, who examine fairness in risk predictions of long-term unemployment, and \citet{körtner:2023}, who study value fairness by incorporating decision maker's inequality aversion into the optimization. This paper emphasizes action fairness and interpretability. 

\defcitealias{Knaus:2020}{Lechner, Knaus, Huber, Frölich, Behncke, Mellace \& Strittmatter (2020)}  

\subsection{Data and Estimation Strategy}

The publicly available administrative dataset from \citetalias{Knaus:2020}, that has been widely used in previous research \citep[e.g.][]{zezulka:2024, körtner:2023, Knaus:2022b, huber:2017}, is employed in this analysis. The dataset covers individuals aged 24 to 55 who were registered as unemployed in Switzerland in 2003 and includes extensive information on individual characteristics and assignments to training programs. This enables the analysis of employment outcomes and fairness properties of different allocations. A detailed description of the data is provided by \cite{Knaus:2022}.

Program participation is documented across several categories, including vocational courses, computer courses, language courses\footnote{This program type includes courses of local and foreign languages.}, employment programs, and a combined category of job search assistance and personality development programs, which are grouped together due to their similar content. The outcome variable is the number of months in employment during 31 months following the program start, which is the latest outcome available in the data. The study relies on the unconfoundedness assumption for causal identification. Following \citet{Knaus:2022}, the analysis controls for a range of features $(X)$ capturing individuals' socioeconomic characteristics and labor market histories. Due to common support issues reported in \citet{Knaus:2022b}, the analysis is restricted to individuals residing in German-speaking cantons (which cover approximately two thirds of the total population).

A common issue in ALMP evaluations is the flexible assignment of participants to programs by caseworkers. Individuals with stronger labor market prospects may find employment before being assigned and are therefore more likely to be in the control group, potentially introducing selection bias. To address this, the approach of \cite{Lechner:1999}, predicting pseudo-treatment start dates for control group members, is applied. For consistent treatment definitions across participants and non-participants, the analysis is restricted to individuals who remain unemployed at their assigned (pseudo) treatment start date. The sample comprises 64,262 individuals.

The sensitive attributes $(S)$ are defined as indicators based on gender (female/male) and nationality (Swiss/foreign). They are  selected for two reasons. First, Switzerland's State Secretariat for Economic Affairs (SECO), which oversees ALMP, explicitly promotes equal opportunities irrespective of gender or nationality \citep[e.g.,][]{SECO:2024, SECO:2022g, SECO:2019}. Second, prior experience with algorithmic tools underscores the relevance of these attributes. For example, a pilot algorithm in Austria was rapidly discontinued after strong public criticism over concerns about unequal treatment by gender and nationality \citep{Achterhold:2025}.
Following \citet{Knaus:2022b}, the decision-relevant features $(A)$ include age of the jobseeker and earnings prior to unemployment. Unlike \citet{Knaus:2022b}, employability is not included, as it is assessed subjectively by caseworkers. Instead, an indicator for whether the unemployed individual has obtained a degree is added.

The Modified Causal Forest (MCF) estimator of \citet{Lechner:2018} is employed to estimate scores, using default settings.\footnote{MCF version 0.7.1 is used. The only deviation from the default configuration is the efficient version for computing IATEs, in which the roles of the samples used for constructing and populating the forest are reversed and the resulting estimates are subsequently averaged.}
The sample is split into three mutually exclusive subsamples: 40\% (25,705 observations) for training the MCF (sample 1),  40\% (25,704 observations) for predicting scores ($\Gamma_d^{\text{IAPO}}$) and training the policy trees (sample 2), and 20\% (12,853 observations) for evaluating policy tree assignments (sample 3). Although theoretical results suggest $\Gamma^{\text{AIPW}}$ as an alternative, $\Gamma^{\text{IAPO}}$ is preferred since \citet{Hatamyar:2023} show that policy trees based on estimated IAPOs outperform those learned from AIPW scores in several simulation settings.
To address remaining common support issues, the min-max trimming of propensity scores described in \citet{Lechner:2019} is applied, using estimates from a random forest classifier. This excludes 2,529 observations (9.8\%) from sample 1, 1,962 (7.6\%) from sample 2, and 993 (7.7\%) from sample 3. A detailed overview of all sample selection steps and resulting sample sizes is provided in Appendix Table \ref{tab:sample:size}.

\subsection{Empirical Results}
\label{52-emp-res}

\begin{figure}[!ht]
    \centering
    \captionsetup{font=small, width=\linewidth} 
    \caption{Distributions of decision-relevant features and one score before and after fairness-adjustment}
    \includegraphics[width=\linewidth]{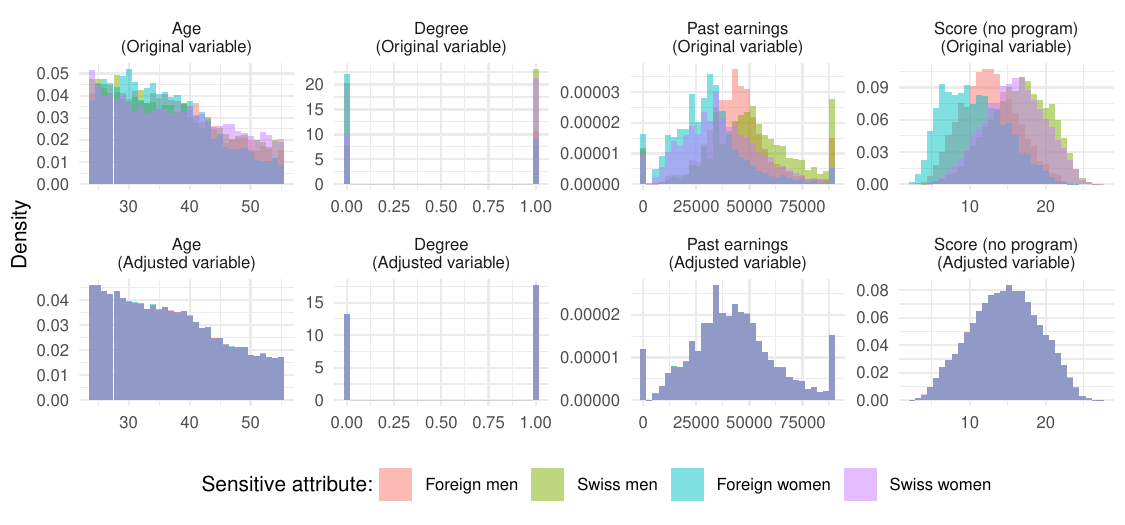}
    \caption*{\scriptsize \textit{Notes:} Histograms of the decision-relevant features (age, degree, past earnings) and of the score for the control group (no program), stratified by sensitive attributes. The top row shows original distributions; the bottom row shows distributions after fairness-adjustment. Score distributions for the five treatment groups can be found in Appendix Figure \ref{fig:histograms_cleaning_appendix}.}    \label{fig:histograms_cleaning_depth_3}
\end{figure}

All results are also available as a notebook on the project webpage.\footnote{\href{https://fmuny.github.io/fairpolicytree/replication_paper/replication_notebook.html}{https://fmuny.github.io/fairpolicytree/replication\_paper/replication\_notebook.html}} Figure \ref{fig:histograms_cleaning_depth_3} shows the distributions of the decision-relevant features and one score, stratified by the four sensitive groups. The top row displays the original distributions, and the bottom row shows the distributions after the \emph{MQ-adjustment} described in Section \ref{attaining_action_fariness}. The figure captures decision-relevant features of varying types, including a binary indicator (degree), and a continuous variable with mass points (past earnings). Before fairness-adjustment, the decision-relevant features vary noticeably across groups. After transformation, the distributions appear closely aligned, suggesting that \emph{MQ-adjustment} effectively mitigated group disparities.

Table \ref{tab:tab:main_results_sample 3} presents the main results comparing policy strategies in terms of interpretability, policy value, fairness, and program allocations. The first two columns list the policy type and whether the policy is interpretable. The third column reports the policy value, defined as the mean potential outcome under the given policy. Columns four to six show three fairness metrics: (i) Cramér's V, a normalized measure of association between categorical variables, ranging from 0 (no association) to 1 (perfect association), (ii) the $p$-value of the associated $\chi^2$-statistic in Cramér's V, and (iii) the logarithm of the Bayes Factor (log(BF)), which quantifies evidence for independence in treatment allocations across sensitive groups (negative values) against the evidence for dependence (positive values). Detailed descriptions of fairness metrics are in Appendix \ref{app:fair}. The final six columns show the program shares, i.e., the proportions of individuals assigned to each intervention under the respective policy.

\renewcommand{\arraystretch}{1.2}
\setlength{\tabcolsep}{3pt}
\begin{table}[t]
\centering
\caption{\label{tab:tab:main_results_sample 3} Policy Value, Fairness and Interpretability for different policies (sample 3)}
\centering
\fontsize{8}{10}\selectfont
\begin{threeparttable}
\begin{tabular}[t]{llrrrrrrrrrr}
\toprule
\multirow{2}{*}{\textbf{Policy}} & \multirow{2}{*}{\makecell{\textbf{Inter-}\\ \textbf{pret.}}} &  \multirow{2}{*}{\makecell{\textbf{Policy}\\ \textbf{value}}} & \multicolumn{3}{c}{\textbf{Fairness}} & \multicolumn{6}{c}{\textbf{Program shares}} \\
\cmidrule(l{3pt}r{3pt}){4-6} \cmidrule(l{3pt}r{3pt}){7-12}
& &  & Cram.V & p-val. & log(BF) & NP & JS & VC & CC & LC & EP\\
\midrule
\addlinespace[0.3em]
\multicolumn{12}{l}{\textbf{Benchmark policies}}\\
\hspace{1em}Observed & False & 14.528 & 0.065 & 0.000 & 18 & 74.5\% & 19.7\% & 1.3\% & 1.4\% & 2.1\% & 0.9\%\\
\hspace{1em}Blackbox & False & 18.174 & 0.097 & 0.000 & 117 & 0.0\% & 0.0\% & 36.9\% & 62.9\% & 0.2\% & 0.0\%\\
\hspace{1em}Blackbox fair & False & 18.152 & 0.028 & 0.001 & -27 & 0.0\% & 0.0\% & 37.5\% & 62.0\% & 0.4\% & 0.1\%\\
\hspace{1em}All in one & True & 17.873 & 0.000 & 1.000 & -Inf & 0.0\% & 0.0\% & 0.0\% & 100.0\% & 0.0\% & 0.0\%\\
\addlinespace[0.3em]
\multicolumn{12}{l}{\textbf{Optimal policy tree (depth 3)}}\\
\hspace{1em}Unadjusted incl.~$S$ & True & 17.909 & 0.459 & 0.000 & 1199 & 0.0\% & 0.0\% & 20.1\% & 79.9\% & 0.0\% & 0.0\%\\
\hspace{1em}Unadjusted excl.~$S$ & True & 17.882 & 0.266 & 0.000 & 395 & 0.0\% & 0.0\% & 11.5\% & 88.5\% & 0.0\% & 0.0\%\\
\hspace{1em}Adjust $A$ & False & 17.883 & 0.089 & 0.000 & 35 & 0.0\% & 0.0\% & 12.4\% & 87.6\% & 0.0\% & 0.0\%\\
\hspace{1em}Adjust $\Gamma_{d}$ & True & 17.880 & 0.248 & 0.000 & 359 & 0.0\% & 0.0\% & 14.5\% & 85.5\% & 0.0\% & 0.0\%\\
\hspace{1em}Adjust $A$ and $\Gamma_{d}$ & False & 17.883 & 0.090 & 0.000 & 36 & 0.0\% & 0.0\% & 18.1\% & 81.9\% & 0.0\% & 0.0\%\\
\addlinespace[0.3em]
\multicolumn{12}{l}{\textbf{Probabilistic split tree (depth 3)}}\\
\hspace{1em}Adjust $A$ & True & 17.884 & 0.087 & 0.000 & 33 & 0.0\% & 0.0\% & 13.0\% & 87.0\% & 0.0\% & 0.0\%\\
\hspace{1em}Adjust $A$ and $\Gamma_{d}$ & True & 17.882 & 0.077 & 0.000 & 23 & 0.0\% & 0.0\% & 19.2\% & 80.8\% & 0.0\% & 0.0\%\\
\bottomrule
\end{tabular}
\begin{tablenotes}[para]
\item \setstretch{1}\scriptsize \textit{Notes:} This table presents measures of interpretability, policy value, and fairness for various policies. The column \textit{Interpret.}~indicates whether the policy is interpretable. The column \textit{Policy value} reports the mean potential outcome under the respective policy. The next three columns display fairness metrics: Cramer's V, the p-value of its associated $\chi^2$-statistic, and the logarithm of the Bayes Factor. The remaining columns report the program shares under each policy, with \textit{NP} = No Program, \textit{JS} = Job Search, \textit{VC} = Vocational Course, \textit{CC} = Computer Course, \textit{LC} = Language Course, \textit{EP} = Employment Program. All statistics are computed out-of-sample, on data not used for estimating scores or training the policy trees.
\end{tablenotes}
\end{threeparttable}
\end{table}

\renewcommand{\arraystretch}{1.0}

The top four rows in Table \ref{tab:tab:main_results_sample 3} report benchmark policies. The \textit{Observed} policy, corresponding to caseworker program assignments, yields the lowest policy value (14.53) and shows moderate group disparities (Cramér's~V = 0.065), leaving room for improvement.  The \textit{Blackbox} policy maximizes policy value without any fairness-adjustments by assigning individuals to the program associated with their highest estimated score. Although this approach yields the highest policy value (18.17), it lacks interpretability, as it is difficult to understand which values of the features lead to the resulting assignments. Most individuals under this policy are assigned to computer courses (63\%). Fairness can be incorporated into the Blackbox algorithm by adjusting scores in a pairwise manner to promote independence from sensitive attributes. If jointly independent scores were obtainable, this would yield an allocation with the maximum level of fairness. Thus, this approach sets an approximate upper bound on achievable fairness-aware policy value. The resulting \textit{Blackbox fair} policy improves fairness (Cramér's~V = 0.028) with only a negligible drop in policy value (18.15), but it remains uninterpretable. The \textit{All in one} policies assign all individuals to the same program. Among these, assigning everyone to a computer course yields the highest policy value (17.87). This fully interpretable strategy achieves perfect fairness but at the cost of a lower policy value. Hence, this serves as a lower bound relative to the \textit{Blackbox fair} policy, which is fair but uninterpretable. The objective is to find a policy that balances fairness, interpretability, and policy value between these bounds.

Next, policy trees with various inputs are considered.\footnote{For computational efficiency, the number of evaluation points for continuous decision-relevant variables is set to 100.} Two policy trees without fairness-adjustments serve as a starting point. The policy  \textit{Unadjusted incl.~$S$}, optimized using the original scores and decision-relevant features including the two sensitive attributes, is interpretable and achieves high policy value (17.91), but fairness is poor (Cramér's~V = 0.459). A natural first step to improve fairness is to exclude the sensitive attributes $S$ from the set of decision-relevant features. As shown in the row \textit{Unadjusted excl.~$S$}, this leads to a slight reduction in policy value (17.88) due to the loss of information relevant for assignment decisions, but fairness improves substantially (Cramér's~V = 0.266). Despite this improvement, a notable degree of dependence between allocated treatments and sensitive attributes remains. Applying the \emph{MQ-adjustment} to the decision-relevant features reduces this dependence (Cramér's~V = 0.089) while leaving policy value essentially unchanged (17.88), but it sacrifices variable interpretability in the original features. Adjusting only the scores $\Gamma_d$ maintains interpretability but yields only moderate fairness gains (Cramér's~V = 0.248) at a similar policy value. Combining \emph{MQ-adjustments} of $A$ and $\Gamma_d$ does not improve fairness (Cramér's~V = 0.090) relative to \emph{MQ-adjustment} of $A$ alone, with policy value again nearly unchanged (17.88). Across these policies, most individuals are assigned to computer courses (approx.~$85\%$). This allocation seems reasonable given the 2003 context, when computer literacy was increasingly important and  demand for related qualifications was high, in addition to potential longer lock-in effects of other programs.\footnote{All simulated policies assign 100\% of individuals to programs, whereas only 25.5\% are assigned in the observed data due to budget and capacity constraints. These constraints are not considered here to isolate the role of fairness and interpretability. Their incorporation is left to future research.}

The final group of tree policies applies the \emph{CQ-adjustment} in combination with the concept of probabilistic split to regain interpretability of \emph{MQ-adjusted} trees based on $\tilde A$. This transformation introduces only marginal differences in policy value and fairness outcomes due to the probabilistic splits in the nodes. As discussed earlier, adjusting $A$ alone yields reasonable policy value performance (17.88) and a high level of fairness (Cramér's~V = 0.087), with most individuals assigned to computer courses (87\%). When both $A$ and $\Gamma_d$ are adjusted, fairness improves further (Cramér's~V = 0.077), accompanied by a modest decrease in computer course assignments (80.8\%) and an increase in vocational training assignments (19.2\%). The level of policy value is of similar size as for the other policy trees.

To summarize, fairness-aware Blackbox policy maximizes policy value under fairness considerations but lacks interpretability, while all-in-one policies are fully fair and interpretable but  reduce policy value. Policy trees with fairness-adjustments and probabilistic splits offer a middle ground, balancing policy value, fairness, and interpretability. In this application, the probabilistic split tree with adjustment of $A$ and $\Gamma_d$ provides the most favorable alignment of these objectives.

\begin{figure}[ht!]
    \centering
    \captionsetup{font=small, width=\linewidth} 
    \caption{Probabilistic split tree}
    \includegraphics[width=0.97\linewidth]{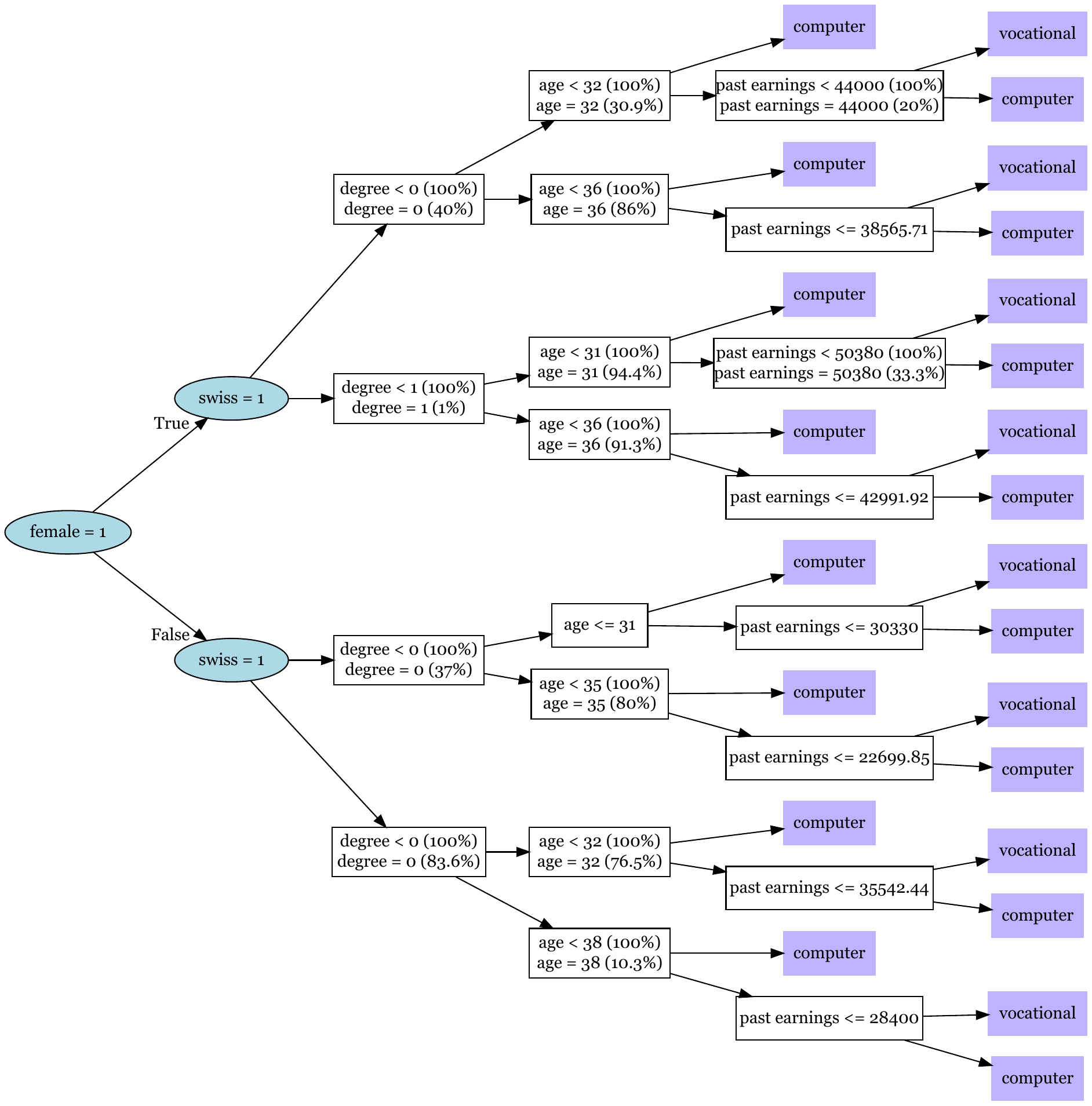}
    \caption*{\scriptsize \textit{Notes:} 
    This figure shows a depth-3 probabilistic policy tree, with splitting thresholds derived from the data. The blue ovals represent deterministic first splits by sensitive groups. Subsequent splits are based on the variables and thresholds reported in white rectangles. At nodes labeled with percentages, the indicated share of individuals whose feature value equals the splitting threshold follows the upper branch; the rest follows the lower branch. Terminal nodes, shown in violet, indicate the allocated program.}    \label{fig:probtree_extleft_depth_3}
\end{figure}

Figure~\ref{fig:probtree_extleft_depth_3} provides a visual representation of the preferred probabilistic split tree based on adjusted $A$ and $\Gamma_d$. The first two splits are determined by sensitive attributes, as the \emph{CQ-adjustment} constructs a separate policy tree for each sensitive group. Subsequent splits separate individuals by degree status, followed by age and past earnings, with splitting thresholds which differ across the sensitive groups. 
The group-specific thresholds are necessary to ensure fairness, offering equal opportunity to treatments across sensitive groups. Ultimately, individuals are  assigned to one of two programs: vocational training or computer training. Individuals under  age 31 are generally assigned to computer training. Older individuals are assigned to computer courses only if their past earnings exceed a certain threshold. Otherwise, they are assigned to vocational training. While age thresholds are relatively similar across sensitive groups, degree and past earnings thresholds differ substantially. The sharpest contrast is between Swiss men and non-Swiss women, likely reflecting  differences in qualifications and earnings.

\begin{figure}[t]
    \centering
    \captionsetup{font=small, width=0.95\linewidth} 
    \caption{Partial fairness-adjustment by linear interpolation}
    \includegraphics[width=0.95\linewidth]{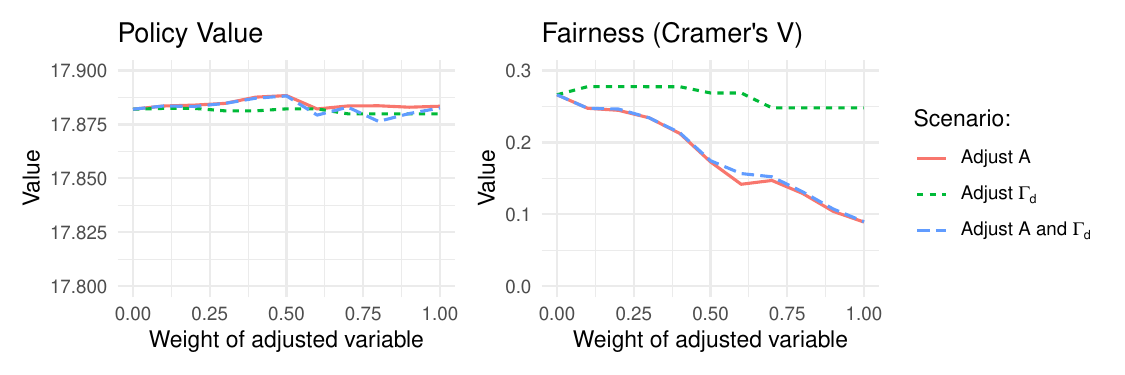}
    \caption*{\scriptsize \textit{Notes:} This figure illustrates the impact of partial fairness-adjustments using linear interpolations between unadjusted and fully adjusted variables as inputs to a depth-3 policy tree. The left panel shows policy value, and the right panel displays fairness (measured as Cramér's V) as a function of the weight of the adjusted variable.}    \label{fig:partial_adjustment}
\end{figure}

Comparison of assignments based on either the original or fully adjusted variables can be extended by considering partially adjusted variables \citep{Feldman:2015}. Gradually increasing the weight of the adjusted variable can provide a more nuanced understanding of the interaction between policy value and fairness. Specifically, define
\begin{align*}
\check{A} = (1 - \lambda) A + \lambda \tilde{A} \quad\quad \text{and} \quad\quad \check{\Gamma}_d = (1 - \lambda) \Gamma_d + \lambda \tilde{\Gamma}_d\text{,}
\end{align*}
where $\lambda \in [0, 1]$ governs the relative weight of the components. The weighted variables $\check{A}$ and/or $\check{\Gamma}_d$ can then be used as inputs for policy trees. Figure \ref{fig:partial_adjustment} illustrates the results over a grid of $\lambda$'s. The left panel shows that policy value remains almost constant regardless of the weight. In contrast, the right panel displays a clear improvement in fairness, as measured by Cramér's V, when adjusting $A$ alone or both $A$ and $\Gamma_d$. The line representing the adjustment of $\Gamma_d$ alone remains relatively flat, indicating limited fairness gains. This highlights that fairness improvements are primarily driven by adjusting $A$, while policy value is largely unaffected across all scenarios. Hence, in this application,  partial fairness is not needed to further address alignment between  policy value and fairness. In other settings, however, where policy value may drop sharply at some point, partial fairness can provide a better balance between policy value and fairness.

As shown above, fairness-adjustments in the policy trees do not substantially change the policy value. Nevertheless, some groups of individuals gain or lose from the enforcement of fairness constraints in interpretable policies. To better understand the characteristics of those benefiting or losing by reassignemnts, $K$-means++ clustering\footnote{$K$-means clustering is an unsupervised machine learning technique that partitions data into $K$ distinct clusters by minimizing the variance within each cluster. The number of clusters is determined in a data-driven way based on the Silhouette score.} \citep{Arthur:2007} is applied to group individuals according to the difference in their individual program scores under the fairness-unaware policy tree (excluding $S$) and the fairness-aware policy tree (with adjusted $A$ and $\Gamma_d$) from Figure~\ref{fig:probtree_extleft_depth_3}.

Table \ref{tab:tab:kmeans_short_pst_adjust_A_score} presents mean values of selected features across the identified clusters.\footnote{Cluster means for all features are provided in Appendix Table~\ref{tab:tab:kmeans_long_pst_adjust_A_score}.} The clustering algorithm distinguishes five groups: two groups of individuals who lose from the reassignment, two groups who benefit, and one group unaffected by the policy change on average. Specifically, 577 individuals (4.9\%) experience a loss in their program score, while another 595 individuals (5.0\%) experience an improvement. The remaining 90.1\% of individuals do not experience a change on average.
Compared to those who lose, individuals who benefit tend to have more stable employment histories (fewer prior unemployment spells), and are less likely to reside in large cities. 
Differences between affected individuals, whether positively or negatively, and unaffected individuals are particularly evident in the decision-relevant features.
Affected individuals are older, have lower prior earnings, and are less likely to hold formal qualifications, indicating greater labor market vulnerability. These patterns suggest that the policy change primarily reallocates treatments within groups with weaker labor market attachment, rather than shifting it between groups with stronger and weaker labor market attachment. Overall, the clustering analysis proves to be a valuable tool for assessing the impact of fairness-adjustments across population subgroups.\footnote{Note that the results of the clustering analysis depend on the chosen benchmark policy. For example, taking the fairness-unaware \textit{Blackbox} policy instead of the fairness-unaware policy tree (excl.~$S$) as a benchmark for the fairness-aware policy tree (adjusted $A$ and $\Gamma_d$) would yield either groups of individuals whose treatment assignment remains unchanged or individuals who lost due to treatment reassignment. There would be no one who gains as \textit{Blackbox} assigns everyone to their best program.}

\renewcommand{\arraystretch}{1.2}
\begin{table}[t]
\centering
\caption{\label{tab:tab:kmeans_short_pst_adjust_A_score}Covariate means of winners and losers from fairness-based reassignment}
\centering
\fontsize{8}{10}\selectfont
\begin{threeparttable}
\begin{tabular}[t]{l>{\raggedleft\arraybackslash}p{1.5cm}>{\raggedleft\arraybackslash}p{1.5cm}>{\raggedleft\arraybackslash}p{1.5cm}>{\raggedleft\arraybackslash}p{1.5cm}>{\raggedleft\arraybackslash}p{1.5cm}}
\toprule
\multicolumn{1}{c}{\textbf{ }} & \multicolumn{5}{c}{\textbf{Cluster (sorted by policy value change)}} \\
\cmidrule(l{3pt}r{3pt}){2-6}
Variable & Strong Loss & Moderate Loss & No Change & Moderate Gain & Strong Gain\\
\midrule
Difference in policy value & -2.16 & -1.00 & 0.00 & 0.93 & 1.97\\
Number of observations in cluster & 185 & 392 & 10688 & 399 & 196\\
\midrule
Job seeker is female & 0.54 & 0.58 & 0.43 & 0.52 & 0.58\\
Swiss citizen & 0.79 & 0.59 & 0.64 & 0.65 & 0.73\\
\midrule
Age of job seeker & 42.41 & 42.32 & 36.00 & 42.18 & 42.57\\
Earnings in CHF before unemployment & 25671 & 27756 & 43965 & 27246 & 28049\\
Qualification: with degree & 0.62 & 0.42 & 0.58 & 0.47 & 0.43\\
\midrule
Fraction months employed in last 2 years & 0.72 & 0.76 & 0.81 & 0.78 & 0.81\\
Employment spells in last 5 years & 1.23 & 1.20 & 1.19 & 1.11 & 0.93\\
Job seeker is married & 0.50 & 0.61 & 0.46 & 0.63 & 0.64\\
Mother tongue other than German, French, Italian & 0.15 & 0.42 & 0.33 & 0.40 & 0.34\\
Unemployment spells in last 2 years & 0.62 & 0.72 & 0.56 & 0.62 & 0.43\\
Lives in no city & 0.56 & 0.65 & 0.68 & 0.75 & 0.77\\
Lives in medium city & 0.14 & 0.11 & 0.13 & 0.14 & 0.14\\
Lives in big city & 0.30 & 0.24 & 0.20 & 0.12 & 0.10\\
\bottomrule
\end{tabular}
\begin{tablenotes}[para]
\item \setstretch{1}\scriptsize \textit{Notes:} The table shows mean values of variables within clusters obtained via $K$-means++ clustering. Clustering is based on the difference in policy value between fairness-aware policy tree (adjusted $A$ and $\Gamma_d$) and fairness-unaware policy tree (excl.~$S$). The number of clusters is determined in a data-driven way by the Silhouette score and the minimum required cluster size is set to 1\% of the observations.
\end{tablenotes}
\end{threeparttable}
\end{table}

\renewcommand{\arraystretch}{1}

The full 31-month post-program period includes the lock-in effect, whereby individuals may temporarily remain unemployed due to program participation itself (i.e., being unavailable for work while attending a program). To better capture long-lasting impacts, Appendix~\ref{app:outcomes} presents additional results for employment outcomes in months 13-24 (second year) and 20-31 (final year available), which exclude the lock-in phase. These alternative specifications offer additional insights into both the empirical analysis and the methodology. For the two additional outcome variables, fairness-adjustments result in a slightly greater loss of policy value relative to the main results. Program assignments also become more evenly distributed between vocational and computer courses, suggesting a stronger lock-in effect for the former. Fairness also improves substantially. Cramér's V using the fairness-adjusted methods approaches zero, and statistical independence between assigned treatments and sensitive attributes can no longer be rejected. This indicates that even pairwise (rather than joint) adjustments to decision-relevant variables can effectively improve fairness.

Another noteworthy finding is that probabilistic split trees may introduce fairness trade-offs. For example, in Table~\ref{tab:tab:main_results_ou_2031}, which evaluates outcomes in the final year, Cramér's V increases from 0.008 for the uninterpretable tree (adjusted $A$) to 0.028 for the interpretable probabilistic split tree (adjusted $A$), indicating that the added randomness in the splits to regain interpretability may come at a fairness cost. A further insight emerges when comparing fairness-unaware policy trees. For both outcomes, including the sensitive attribute $S$ in the decision-relevant features results in fairer allocations than excluding it, which initially seems paradoxical. As noted above, omitting sensitive attributes does not necessarily improve fairness, as other variables can act as proxies. In this application, these proxies appear to even increase unfairness relative to including $S$ directly. In contrast, the proposed fairness-adjustments consistently lead to strong improvement in terms of fairness.

\section{Conclusion}\label{conclusion}

This paper addresses the challenge of combining fairness and interpretability in algorithmic decision making by proposing an extended framework for fairness-aware policy learning. The suggested approach is based on a pre-processing step that removes dependencies between sensitive and decision-relevant variables, combined with a method to retain the interpretability of policy trees by translating splitting rules back to the original variable space. This ensures that fairness in the resulting allocations can be improved without sacrificing interpretability, an essential requirement for potential adoption. 
Applied to Swiss unemployment data, the method shows that fair and interpretable treatment allocations can be achieved with only a minimal reduction in policy value compared to policy value-maximizing but potentially unfair allocations.
The reallocation of individuals from interpretable to interpretable and fair allocations results in a loss for some individuals, while others benefit from the change. Notably, these effects occur primarily within groups with weaker labor market attachment, meaning the policy shift redistributes resources among members of these groups, rather than between individuals with weaker and stronger labor market attachment.

While the suggested approach demonstrates promising results, some avenues remain open for future research. First, extending the procedure to enforce joint independence, rather than only pairwise independence, between sensitive attributes and decision-relevant variables could lead to stronger fairness guarantees and warrants further investigation, especially in relation to interpretability. Second, extending the method to incorporate broader fairness criteria that permit indirect influence of sensitive attributes through materially-relevant variables \citep{Strack:2023} could further enhance the applicability of the framework. Finally, it would be interesting to see additional applications of the policy value-fairness-interpretability 
analysis as a tool to evaluate algorithmic decision tools. 

\newpage

\bibliographystyle{apacite}
\bibliography{library}

\newpage
\begin{appendices}
\renewcommand\thetable{\thesection.\arabic{table}} 
\counterwithin{table}{section}
\renewcommand\thefigure{\thesection.\arabic{figure}} 
\counterwithin{figure}{section}

\section{Additional Results}
\label{Appendix_A}

\subsection{Sample Sizes}
\begin{table}[ht]
\centering
\captionsetup{font=small, width=0.54\linewidth} 
\caption{\label{tab:sample:size}Sample sizes}
\centering
\small
\begin{tabular}{l r}
\toprule
\textbf{Sample Description} & \textbf{Sample size} \\
\midrule
Full sample                 & 100,120 \\
Keep German-speaking cantons only            & 69,372 \\
Correction for pseudo-start dates          & 64,262 \\
\midrule
\multicolumn{2}{l}{\textbf{Sample splitting:}} \\
\quad Sample 1 (train MCF)              & 25,704 \\
\quad Sample 2 (predict MCF, train policy)             & 25,705 \\
\quad Sample 3 (predict policy)             & 12,853 \\
\midrule
\multicolumn{2}{l}{\textbf{Trimming:}} \\
\quad Sample 1  (train MCF)               & 23,175 \\
\quad Sample 2  (predict MCF, train policy)             & 23,742 \\
\quad Sample 3  (predict policy)            & 11,860 \\
\bottomrule
\end{tabular}
\caption*{\scriptsize \textit{Notes:} Table shows sample sizes at different stages of the analysis.}  
\label{tab:sample_sizes}
\end{table}

\subsection{Adjustments of Remaining Scores}
\begin{figure}[H]
    \centering
    \captionsetup{font=small, width=\linewidth} 
    \caption{Distributions of scores before and after fairness-adjustment}
    \includegraphics[width=\linewidth]{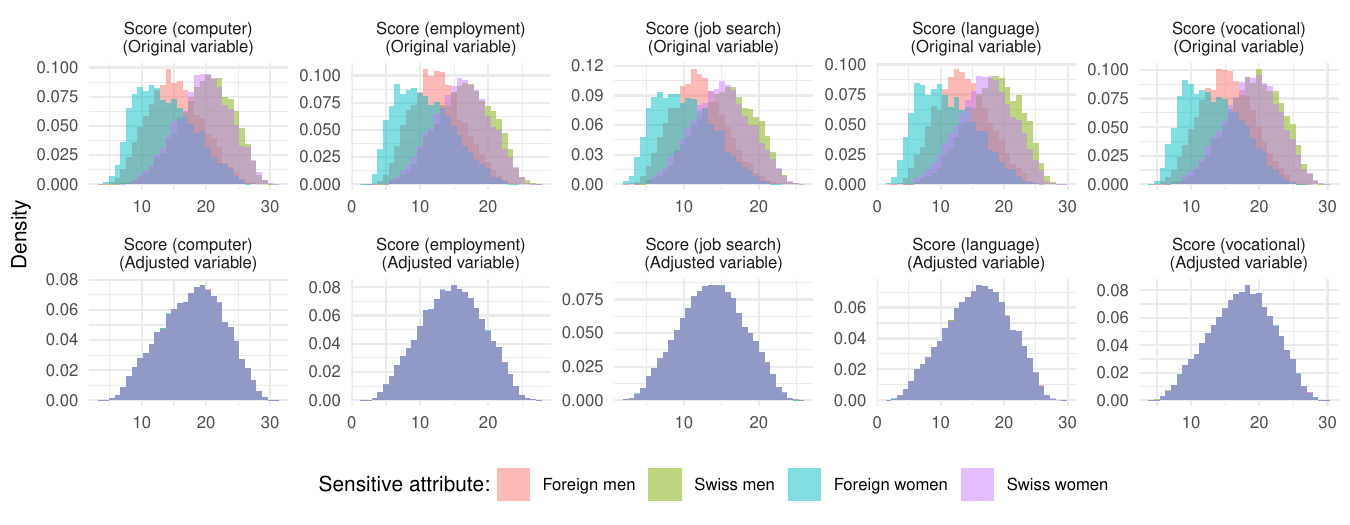}
    \caption*{\scriptsize \textit{Notes:} Score histograms of the five treatment groups, stratified by sensitive attributes. The top row shows original distributions; the bottom row shows distributions after fairness-adjustment. The control group is shown in Figure \ref{fig:histograms_cleaning_depth_3}.}
    \label{fig:histograms_cleaning_appendix}
\end{figure}

\subsection{\textit{K}-means Clustering}
\renewcommand{\arraystretch}{0.7}
\begin{table}[H]
\centering
\caption{\label{tab:tab:kmeans_long_pst_adjust_A_score}Covariate means of winners and losers from fairness-based reassignment (all covariates)}
\centering
\fontsize{8}{10}\selectfont
\begin{threeparttable}
\begin{tabular}[t]{l>{\raggedleft\arraybackslash}p{1.5cm}>{\raggedleft\arraybackslash}p{1.5cm}>{\raggedleft\arraybackslash}p{1.5cm}>{\raggedleft\arraybackslash}p{1.5cm}>{\raggedleft\arraybackslash}p{1.5cm}}
\toprule
\multicolumn{1}{c}{\textbf{ }} & \multicolumn{5}{c}{\textbf{Cluster (sorted by welfare change)}} \\
\cmidrule(l{3pt}r{3pt}){2-6}
Variable & Strong Loss & Moderate Loss & No Change & Moderate Gain & Strong Gain\\
\midrule
Difference in policy value & -2.16 & -1.00 & -0.00 & 0.93 & 1.97\\
Number of observations in cluster & 185 & 392 & 10688 & 399 & 196\\
\midrule
Job seeker is female & 0.54 & 0.58 & 0.43 & 0.52 & 0.58\\
Swiss citizen & 0.79 & 0.59 & 0.64 & 0.65 & 0.73\\
\midrule
Age of job seeker & 42.41 & 42.32 & 36.00 & 42.18 & 42.57\\
Earnings in CHF before unemployment & 25671 & 27756 & 43965 & 27246 & 28049\\
Qualification: with degree & 0.62 & 0.42 & 0.58 & 0.47 & \vphantom{1} 0.43\\
\midrule
Fraction months employed in last 2 years & 0.72 & 0.76 & 0.81 & 0.78 & 0.81\\
Employment spells in last 5 years & 1.23 & 1.20 & 1.19 & 1.11 & 0.93\\
Job seeker is married & 0.50 & 0.61 & 0.46 & 0.63 & 0.64\\
Mother tongue other than German, French, Italian & 0.15 & 0.42 & 0.33 & 0.40 & 0.34\\
Unemployment spells in last 2 years & 0.62 & 0.72 & 0.56 & 0.62 & 0.43\\
Lives in no city & 0.56 & 0.65 & 0.68 & 0.75 & 0.77\\
Lives in medium city & 0.14 & 0.11 & 0.13 & 0.14 & 0.14\\
Lives in big city & 0.30 & 0.24 & 0.20 & 0.12 & 0.10\\
Mother tongue in canton's language & 0.03 & 0.09 & 0.11 & 0.09 & 0.07\\
Age of caseworker & 42.74 & 43.20 & 44.08 & 46.03 & 47.41\\
Caseworker cooperative & 0.59 & 0.51 & 0.49 & 0.39 & 0.32\\
Caseworker education: above vocational training & 0.33 & 0.44 & 0.46 & 0.50 & 0.58\\
Caseworker education: tertiary track & 0.38 & 0.19 & 0.20 & 0.13 & 0.11\\
Caseworker female & 0.54 & 0.49 & 0.44 & 0.43 & 0.35\\
Indicator for missing caseworker characteristics & 0.05 & 0.05 & 0.05 & 0.04 & 0.08\\
Caseworker has own unemployemnt experience & 0.70 & 0.59 & 0.62 & 0.64 & 0.71\\
Caseworker job tenure in years & 4.61 & 5.33 & 5.53 & 5.97 & 6.13\\
Caseworker education: vocational training degree & 0.24 & 0.28 & 0.26 & 0.26 & 0.14\\
Employability as assessed by the caseworker & 1.85 & 1.84 & 1.93 & 1.75 & 1.84\\
Foreigner with temporary permit (B permit) & 0.06 & 0.16 & 0.13 & 0.12 & 0.07\\
Foreigner with permanent permit (C permit) & 0.14 & 0.25 & 0.23 & 0.23 & 0.19\\
Cantonal GDP per capita (in CHF 10,000) & 0.59 & 0.55 & 0.52 & 0.48 & 0.48\\
Allocation to caseworker: by industry & 0.68 & 0.58 & 0.59 & 0.54 & 0.47\\
Allocation to caseworker: by occupation & 0.58 & 0.53 & 0.51 & 0.42 & 0.31\\
Allocation to caseworker: by age & 0.06 & 0.05 & 0.04 & 0.05 & 0.02\\
Allocation to caseworker: by employability & 0.06 & 0.09 & 0.08 & 0.10 & 0.05\\
Allocation to caseworker: by region & 0.08 & 0.16 & 0.12 & 0.13 & 0.16\\
Allocation to caseworker: other & 0.10 & 0.11 & 0.09 & 0.10 & 0.07\\
Cantonal unemployment rate (in \%) & 4.25 & 3.72 & 3.51 & 3.19 & 3.20\\
Sector of last job: tertiary sector & 0.66 & 0.56 & 0.61 & 0.59 & 0.61\\
Sector of last job: secondary sector & 0.07 & 0.12 & 0.14 & 0.13 & 0.12\\
Sector of last job: missing sector & 0.24 & 0.26 & 0.17 & 0.20 & 0.20\\
Sector of last job: primary sector & 0.03 & 0.06 & 0.09 & 0.08 & 0.07\\
Previous job: skilled worker & 0.66 & 0.50 & 0.61 & 0.45 & 0.52\\
Previous job: manager & 0.04 & 0.04 & 0.07 & 0.03 & 0.05\\
Previous job: unskilled worker & 0.24 & 0.44 & 0.28 & 0.50 & 0.41\\
Previous job: self-employed & 0.06 & 0.02 & 0.03 & 0.02 & 0.02\\
Qualification: with degree & 0.62 & 0.42 & 0.58 & 0.47 & 0.43\\
Qualification: semiskilled & 0.09 & 0.18 & 0.16 & 0.20 & 0.32\\
Qualification: unskilled & 0.26 & 0.36 & 0.23 & 0.30 & 0.21\\
Qualification: no degree & 0.02 & 0.03 & 0.04 & 0.03 & 0.04\\
\bottomrule
\end{tabular}
\begin{tablenotes}[para]
\item \setstretch{1}\scriptsize \textit{Notes:} The table shows mean values of variables within clusters obtained via $K$-means++ clustering. Clustering is based on the difference in welfare between optimal policy trees with fairness-adjustment (of decision-relevant features and scores) and without adjustments (excl.~$S$). The number of clusters is determined in a data-driven way by the Silhouette score and the minimum required cluster size is set to 1\% of the observations.
\end{tablenotes}
\end{threeparttable}
\end{table}

\renewcommand{\arraystretch}{1}

\subsection{Results for Additional Outcomes}\label{app:outcomes}

\begin{table}[H]
\centering
\caption{\label{tab:tab:main_results_ou_1324}Main results (sample 3) using alternative outcome variable: Total number of months employed in the second year available in the data (months 13 to 24 after start of program)}
\centering
\fontsize{8}{10}\selectfont
\begin{threeparttable}
\begin{tabular}[t]{llrrrrrrrrrr}
\toprule
\multirow{2}{*}{\textbf{Policy}} & \multirow{2}{*}{\makecell{\textbf{Inter-}\\[-1ex]\textbf{pret.}}} &  \multirow{2}{*}{\makecell{\textbf{Policy}\\[-1ex]\textbf{value}}} & \multicolumn{3}{c}{\textbf{Fairness}} & \multicolumn{6}{c}{\textbf{Program shares}} \\
\cmidrule(l{3pt}r{3pt}){4-6} \cmidrule(l{3pt}r{3pt}){7-12}
& &  & Cram.V & p-val. & log(BF) & NP & JS & VC & CC & LC & EP\\
\midrule
\addlinespace[0.3em]
\multicolumn{12}{l}{\textbf{Benchmark policies}}\\
\hspace{1em}Observed & False & 6.232 & 0.065 & 0.000 & 18 & 74.5\% & 19.7\% & 1.3\% & 1.4\% & 2.1\% & 0.9\%\\
\hspace{1em}Blackbox & False & 7.921 & 0.076 & 0.000 & 53 & 0.0\% & 0.0\% & 46.4\% & 53.5\% & 0.1\% & 0.0\%\\
\hspace{1em}Blackbox fair & False & 7.914 & 0.021 & 0.065 & -34 & 0.0\% & 0.0\% & 46.1\% & 53.6\% & 0.1\% & 0.1\%\\
\hspace{1em}All in one & True & 7.734 & 0.000 & 1.000 & -Inf & 0.0\% & 0.0\% & 0.0\% & 100.0\% & 0.0\% & 0.0\%\\
\addlinespace[0.3em]
\multicolumn{12}{l}{\textbf{Policy tree (depth 3)}}\\
\hspace{1em}Unadjusted incl.~$S$ & True & 7.795 & 0.263 & 0.000 & 399 & 0.0\% & 0.0\% & 42.7\% & 57.3\% & 0.0\% & 0.0\%\\
\hspace{1em}Unadjusted excl.~$S$ & True & 7.782 & 0.379 & 0.000 & 887 & 0.0\% & 0.0\% & 46.8\% & 53.2\% & 0.0\% & 0.0\%\\
\hspace{1em}Adjust $A$ & False & 7.776 & 0.017 & 0.313 & -9 & 0.0\% & 0.0\% & 41.4\% & 58.6\% & 0.0\% & 0.0\%\\
\hspace{1em}Adjust $\Gamma_{d}$ & True & 7.775 & 0.367 & 0.000 & 819 & 0.0\% & 0.0\% & 45.8\% & 54.2\% & 0.0\% & 0.0\%\\
\hspace{1em}Adjust $A$ and $\Gamma_{d}$ & False & 7.776 & 0.017 & 0.313 & -9 & 0.0\% & 0.0\% & 41.4\% & 58.6\% & 0.0\% & 0.0\%\\
\addlinespace[0.3em]
\multicolumn{12}{l}{\textbf{Probabilistic split tree (depth 3)}}\\
\hspace{1em}Adjust $A$ & True & 7.778 & 0.024 & 0.083 & -7 & 0.0\% & 0.0\% & 43.1\% & 56.9\% & 0.0\% & 0.0\%\\
\hspace{1em}Adjust $A$ and $\Gamma_{d}$ & True & 7.776 & 0.016 & 0.409 & -9 & 0.0\% & 0.0\% & 42.2\% & 57.8\% & 0.0\% & 0.0\%\\
\bottomrule
\end{tabular}
\begin{tablenotes}[para]
\item \setstretch{1}\scriptsize \textit{Notes:} This table presents measures of interpretability, policy value, and fairness for various policies. The column \textit{Interpret.} indicates whether the policy is interpretable. The column \textit{Policy value} reports the mean potential outcome under the respective policy. The next three columns display fairness metrics: Cramer's V, the p-value of its associated $\chi^2$-statistic, and the logarithm of the Bayes Factor. The remaining columns report the program shares under each policy, with \textit{NP} = No Program, \textit{JS} = Job Search, \textit{VC} = Vocational Course, \textit{CC} = Computer Course, \textit{LC} = Language Course, \textit{EP} = Employment Program. Statistics are computed out-of-sample, on data not used for estimating scores or training the policy tree.
\end{tablenotes}
\end{threeparttable}
\end{table}

\begin{table}[H]
\centering
\caption{\label{tab:tab:main_results_ou_2031}Main results (sample 3) using alternative outcome variable: Total number of months employed in the final year available in the data (months 20 to 31 after start of program)}
\centering
\fontsize{8}{10}\selectfont
\begin{threeparttable}
\begin{tabular}[t]{llrrrrrrrrrr}
\toprule
\multirow{2}{*}{\textbf{Policy}} & \multirow{2}{*}{\makecell{\textbf{Inter-}\\[-1ex]\textbf{pret.}}} &  \multirow{2}{*}{\makecell{\textbf{Policy}\\[-1ex]\textbf{value}}} & \multicolumn{3}{c}{\textbf{Fairness}} & \multicolumn{6}{c}{\textbf{Program shares}} \\
\cmidrule(l{3pt}r{3pt}){4-6} \cmidrule(l{3pt}r{3pt}){7-12}
& &  & Cram.V & p-val. & log(BF) & NP & JS & VC & CC & LC & EP\\
\midrule
\addlinespace[0.3em]
\multicolumn{12}{l}{\textbf{Benchmark policies}}\\
\hspace{1em}Observed & False & 6.563 & 0.065 & 0.000 & 18 & 74.5\% & 19.7\% & 1.3\% & 1.4\% & 2.1\% & 0.9\%\\
\hspace{1em}Blackbox & False & 8.001 & 0.096 & 0.000 & 121 & 0.0\% & 0.0\% & 51.3\% & 48.1\% & 0.3\% & 0.3\%\\
\hspace{1em}Blackbox fair & False & 7.992 & 0.024 & 0.018 & -27 & 0.0\% & 0.0\% & 50.0\% & 48.6\% & 0.5\% & 0.8\%\\
\hspace{1em}All in one & True & 7.756 & 0.000 & 1.000 & -Inf & 0.0\% & 0.0\% & 0.0\% & 100.0\% & 0.0\% & 0.0\%\\
\addlinespace[0.3em]
\multicolumn{12}{l}{\textbf{Policy tree (depth 3)}}\\
\hspace{1em}Unadjusted incl.~$S$ & True & 7.878 & 0.252 & 0.000 & 374 & 0.0\% & 0.0\% & 51.3\% & 48.7\% & 0.0\% & 0.0\%\\
\hspace{1em}Unadjusted excl.~$S$ & True & 7.872 & 0.354 & 0.000 & 779 & 0.0\% & 0.0\% & 55.6\% & 44.4\% & 0.0\% & 0.0\%\\
\hspace{1em}Adjust $A$ & False & 7.863 & 0.008 & 0.857 & -10 & 0.0\% & 0.0\% & 46.8\% & 53.2\% & 0.0\% & 0.0\%\\
\hspace{1em}Adjust $\Gamma_{d}$ & True & 7.867 & 0.369 & 0.000 & 844 & 0.0\% & 0.0\% & 50.9\% & 49.1\% & 0.0\% & 0.0\%\\
\hspace{1em}Adjust $A$ and $\Gamma_{d}$ & False & 7.863 & 0.008 & 0.857 & -10 & 0.0\% & 0.0\% & 46.8\% & 53.2\% & 0.0\% & 0.0\%\\
\addlinespace[0.3em]
\multicolumn{12}{l}{\textbf{Probabilistic split tree (depth 3)}}\\
\hspace{1em}Adjust $A$ & True & 7.860 & 0.028 & 0.028 & -6 & 0.0\% & 0.0\% & 47.5\% & 52.5\% & 0.0\% & 0.0\%\\
\hspace{1em}Adjust $A$ and $\Gamma_{d}$ & True & 7.860 & 0.029 & 0.018 & -6 & 0.0\% & 0.0\% & 48.4\% & 51.6\% & 0.0\% & 0.0\%\\
\bottomrule
\end{tabular}
\begin{tablenotes}[para]
\item \setstretch{1}\scriptsize \textit{Notes:} This table presents measures of interpretability, policy value, and fairness for various policies. The column \textit{Interpret.} indicates whether the policy is interpretable. The column \textit{Policy value} reports the mean potential outcome under the respective policy. The next three columns display fairness metrics: Cramer's V, the p-value of its associated $\chi^2$-statistic, and the logarithm of the Bayes Factor. The remaining columns report the program shares under each policy, with \textit{NP} = No Program, \textit{JS} = Job Search, \textit{VC} = Vocational Course, \textit{CC} = Computer Course, \textit{LC} = Language Course, \textit{EP} = Employment Program. Statistics for computed out-of-sample, on data not used for estimating scores or training the policy tree.
\end{tablenotes}
\end{threeparttable}
\end{table}

\section{Technical Appendix}
\label{Appendix_B}

This appendix contains more details on fairness metrics, Cramér's V and Bayes factor, applied in Section \ref{52-emp-res}.

\subsection{Fairness Metrics}
\label{app:fair}
Cramér's V measures the association between two categorical variables with more than two categories, where Pearson correlation is unsuitable. It is defined as $V =\sqrt{\frac{\chi^2}{N(C-1)}}$  where $\chi^2$ is the test statistic of a chi-squared independence test, $N$ is the total sample size, and $C$ is the smaller number of categories in the two variables. Unlike the raw $\chi^2$ statistic, Cramér's V provides an interpretable effect size ranging from 0 (no association) to 1 (perfect association) and is less sensitive to sample size \citep{Kearney:2017}. 

Bayes factor for the analysis of contingency tables under an independent multinomial sampling scheme following \cite{Gunel:1974} is applied.
It is defined as the ratio of two marginal likelihoods, each quantifying how likely it is to observe the data under a given hypothesis. The null assumes equal multinomial distributions of treatment allocations across sensitive groups, while the alternative captures deviations from this assumption.
The Bayes factor offers a valuable complement to Cramér's V by allowing researchers to quantify evidence for and against the null hypothesis, offering a more balanced alternative to $p$-values, which are known to overestimate evidence against the null hypothesis \citep[e.g.,][]{Berger:1987}. Because some of the evaluated policies yield large Bayes factors, the results are reported on the logarithmic scale (log(BF)). Since the null marginal likelihood appears in the denominator, the log Bayes factor can be interpreted as follows. {A value of 0 indicates no evidence either for dependence or independence. Values greater than 0 suggest evidence for dependence, with values above 2 indicating strong evidence. Conversely, values below 0 suggest independence, with values below -2 representing strong evidence for independence \citep{Jamil:2017}.}

\end{appendices}

\end{document}